\documentclass[a4paper,11pt]{article}

\usepackage{graphicx}
\usepackage{url}
\usepackage{natbib}
\usepackage{fancyhdr}

\def\deg{$^{\circ}$}

\def\oisotope{$^{18}$O}
\def\oproxy{$\delta ^{18}$O}
\def\co2{CO$_2$}
\def\no2{NO$_2$}

\def\omp{$\Omega_p$}
\def\kms{km\,s$^{-1}$}
\def\kmskpc{km\,s$^{-1}$\,kpc$^{-1}$}
\def\kgm{kg\,m$^{-3}$}
\def\uas{$\mu$as}

\begin{document}

%%%%%
\begin{titlepage}

\begin{centerline}
{\LARGE {\bf The evidence for and against astronomical impacts on}}
\end{centerline}
\vspace*{0.4cm}
\begin{centerline}
{\LARGE {\bf climate change and mass extinctions: A review}}
\end{centerline}

\vspace*{1cm}

\begin{centerline}
{\Large Coryn A.L.\ Bailer-Jones}
\end{centerline}
\vspace*{0.5cm}
\begin{centerline}
{Max-Planck-Institut f\"ur Astronomie, K\"onigstuhl 17, Heidelberg, Germany}
\end{centerline}
\vspace*{0.15cm}
\begin{centerline}
{Email: calj@mpia.de}
\end{centerline}

\vspace*{0.8cm}

\begin{centerline}
{to appear in the {\em International Journal of Astrobiology}}
\end{centerline}
\vspace*{0.2cm}
\begin{centerline}
{\small submitted 15 April 2009; accepted 20 May 2009; revised 23 May 2009 and 23 June 2009}
\end{centerline}

\vspace*{1.0cm}

\section*{Abstract}
Numerous studies over the past 30 years have suggested there is a causal connection between the motion of the Sun through the Galaxy and terrestrial mass extinctions or climate change.  Proposed mechanisms include comet impacts (via perturbation of the Oort cloud), cosmic rays and supernovae, the effects of which are modulated by the passage of the Sun through the Galactic midplane or spiral arms.  Supposed periodicities in the fossil record, impact cratering dates or climate proxies over the Phanerozoic (past 545\,Myr) are frequently cited as evidence in support of these hypotheses.  This remains a controversial subject, with many refutations and replies having been published.  Here I review both the mechanisms and the evidence for and against the relevance of astronomical phenomena to climate change and evolution. This necessarily includes a critical assessment of time series analysis techniques and hypothesis testing.  Some of the studies have suffered from flaws in methodology, in particular drawing incorrect conclusions based on ruling out a null hypothesis.  I conclude that there is little evidence for intrinsic periodicities in biodiversity, impact cratering or climate on timescales of tens to hundreds of Myr.  Although this does not rule out the mechanisms, the numerous assumptions and uncertainties involved in the interpretation of the geological data and in particular in the astronomical mechanisms, suggest that Galactic midplane and spiral arm crossings have little impact on biological or climate variation above background level.  Non-periodic impacts and terrestrial mechanisms (volcanism, plate tectonics, sea level changes), possibly occurring simultaneously, remain likely causes of many environmental catastrophes. Internal dynamics of the biosphere may also play a role.  In contrast, there is little evidence supporting the idea that cosmic rays have a significant influence on climate through cloud formation.  It seems likely that more than one mechanism has contributed to biodiversity variations over the past half Gyr.

{\bf Keywords}: {\em mass extinctions, climate change, solar motion, spiral arms, minor body impacts, cosmic rays, hypothesis testing, time series analysis, period detection}

\end{titlepage}
%%%%%

\pagenumbering{arabic}
\setcounter{page}{2}

\newpage
\tableofcontents

%%%%%%%%%%%%%%%%%%%%%%%%%%%%%%%%%%%%%%%%%%%%%%%%%%%%
\section{Introduction}

Do astronomical phenomena have an impact on life on Earth? The answer is of course ``yes''. The seasons are a result of the Earth's orbit around the Sun and the ice ages over the past few hundred thousand years were almost certainly caused by well-understood changes in this orbit and the orientation of the Earth's axis. In this article I will primarily examine changes which took place over a longer timescale, tens or hundreds of millions of years. On these timescales other mechanisms connected to the orbit of the Sun around the Galaxy come into considerations, a priori at least. It is my objective to examine the evidence for and against various astronomical mechanisms for causing mass extinctions and/or climate change.

I start in section~\ref{ts_evidence} by examining the data on variations (in particular periodicities) in the geological and biological records. These include biodiversity, impact cratering and climate proxies.  The analyses in the literature, and criticisms thereof, raise a number of issues concerning the nature of hypothesis testing, which I discuss in more detail in section~\ref{ts_fun}. There I draw attention in particular to the limitation of rejecting a null hypothesis based just on Monte Carlo simulations, a limitation I call ``incomplete inference''. After briefly summarizing possible terrestrial mechanisms (section~\ref{terra_mech}), I describe the various extraterrestrial mechanisms which have been proposed to influence life and/or climate on Earth (section~\ref{ET_mechanisms}). These include minor body impacts, cosmic rays, supernovae and gamma-ray bursts, solar variations and changes in the Earth's orbit. To create variation these mechanisms require triggers, and in section~\ref{solar_motion} I discuss two aspects of the solar motion about the Galaxy which have been proposed: the vertical oscillation about the Galactic plane and spiral arm crossing. In section~\ref{improving} I outline how the upcoming Gaia astrometric survey may be able to improve the situation. I conclude in section~\ref{conclusions}.

A lot has been published on this topic in the past 30 years.  It is not my intention to comprehensively review the literature, but rather to present the major themes. Some earlier reviews on parts of this topic include Torbett~\citep{torbett_1989} and Rampino~\citep{rampino_1998}. A more popular discussion of the causes of mass extinction is Hallam~\citep{hallam_2004} and an interesting history and analysis of the impacts vs.\ volcanism debate is given by Glen~\citep{glen_1994} (chapters 1 and 2).

%%%%%%%%%%%%%%%%%%%%%%%%%%%%%%%%%%%%%%%%%%%%%%%%%%%%
\section{Evidence from geological time series}\label{ts_evidence}

\subsection{Geological and biological data}\label{geobiodata}

There are various types of geological and biological data which are used to study climate history
or evolution. Widely used measures of biological change are as
follows.
\vspace*{-0.5em}
\begin{itemize}
\item Species or genera diversity. This measures the number of different species or genera present at any one time.  A species is the lowest level taxonomic rank; a genus is composed of many species (above this comes family, order, class etc.).
\item Extinction. This can be expressed either as the number (or fraction) of species or genera which become extinct in a time interval, or as a time series of delta functions marking epochs of mass extinction but without any measure of the extent of the extinction.
\item Origination. As with extinction, but for newly created species or genera.
\end{itemize}
The above are measured via the fossil record and have been recorded back to at least the beginning of the Phanerozoic eon, some 545\,Myr BP (before present).  This time marks a significant increase in the diversity of life on Earth, in particular the occurrence of hard-shelled animals. 
Diversity, extinction and origination are not necessarily correlated. For example, a large extinction concurrent with a large origination event would result in little change in the total diversity.

There are several tracers of geological activity
\vspace*{-0.5em}
\begin{itemize}
\item Impact craters.
\item Iridium layers. It has been proposed that iridium -- which is rare in the Earth's crust -- could be delivered by a meteor or comet and then spread over a large area following the impact. This was proposed by Alvarez et al.~\citep{alvarez_etal_1980} as evidence of an impact at the K-T (Cretaceous--Tertiary) boundary 65\,Myr BP.
\item Flood basalts. Giant volcanic eruptions result in basalt lava covering large areas of land or sea crust. As the lava has low viscosity it forms layers rather than a classic volcano, although on giant scales it can form mountain ranges and plateaus. Examples are the Deccan traps in India (occurring around 65\,Myr BP) and the Siberian traps (around 250\,Myr BP).
\item Orogenic events (mountain building) and plate tectonics.  These affect climate because of their influence on atmospheric and oceanic circulation as well as the formation of icecaps at the poles.
\item Geomagnetic reversals, the orientation of the field being preserved in some rocks.
\item \oisotope\ temperature proxy. \oisotope\ is heavier than the much more abundant oxygen isotope $^{16}$O, so evaporation of water leads to a partial separation of these (stable) isotopes.  The degree of separation is temperature dependent: the warmer the water, the higher the \oisotope\ content of the evaporated water.  Hence the relative abundance of \oisotope\ to $^{16}$O (written \oproxy) may be used as proxy for the ocean temperature (Dansgaard~\citealp{dansgaard_1964}).  When this water is precipitated out of the atmosphere as rain or snow it can remain permanently frozen in places such as Greenland and Antarctica.
There is then a positive correlation between \oisotope\ and temperature, e.g.\ \oproxy\,$= 0.7\,T - 13.6$, where \oproxy\ is measured in parts per thousand and $T$ in Celsius (the constants depend on geographical location). This has been recorded in several ice cores going back as far as 5.3\,Myr BP (Lisiecki \& Raymo~\citealp{lisiecki_raymo_2005}). Some of these cores, e.g.\ the Vostok and EPICA ice cores, have also measured \oproxy\ of the air trapped in the ice, thus providing a measure of atmospheric temperatures.  The precipitated water can also be taken up by land animals and \oproxy\ later measured in their fossils,
whereby there is also a positive correlation.  In contrast, the oceans from which the water evaporated become depleted in \oisotope. This can be measured in marine fossils, in particular in the calcite shells of foraminifera (mm sized creatures) on the ocean floor. In this case, \oproxy\ is negatively correlated with the ocean temperature at the time the organism died (Epstein et al.~\citealp{epstein_etal_1961}). Such fossils have been used to trace climate back as far as 600\,Myr BP (e.g.\ Veizer et al.~\citealp{veizer_etal_1999}, IPCC~\citealp{ipcc_2007}).
\item Sea level variations, driven either by local uplift (e.g.\ plate tectonics) or through changes in the amount of water locked in glaciers and ice caps.
\item Anoxic events. These are periods when oxygen is completely depleted from the oceans and so are implicated in mass extinctions. They are believed to be a consequence of changes in climate and ocean circulation.
\end{itemize}

In this section I will review several studies which claim to have detected periodicities in one or more of the above records. Periodicity detection is often controversial because there are many different ways to analyse time series and there are differences in opinion of what constitutes a significant detection. Yet it is appealing, because detection of a period makes it easier to associate extinction events to a (recurring) physical mechanism.  However, periodicities in the data do not automatically imply periodicities in the driver; they may just reflect the characteristic response of the system to an impulsive trigger. 
More generally, 
triggers of extinction could be pulsed but the system may respond smoothly.
On the other hand, a smoothly varying external driver could give rise to a smooth response in the system (e.g.\ species origination and extinction), but due to limited sensitivity (e.g.\ preservation in the fossil record) only those events which rise above some background level are observed. An underlying continuous but variable phenomenon is thus converted into an apparently pulsed one.

With these considerations in mind, let us look at claims of periodicities in the geological data.

%%%%%%%%%%%%%%%%%%%%%%%%%%%%%%%%%%%%%
\subsection{Periods in climate data over the past few million years}\label{climate_myr}

\begin{figure}[htb]
\begin{center}
\includegraphics[scale=0.40, angle=0]{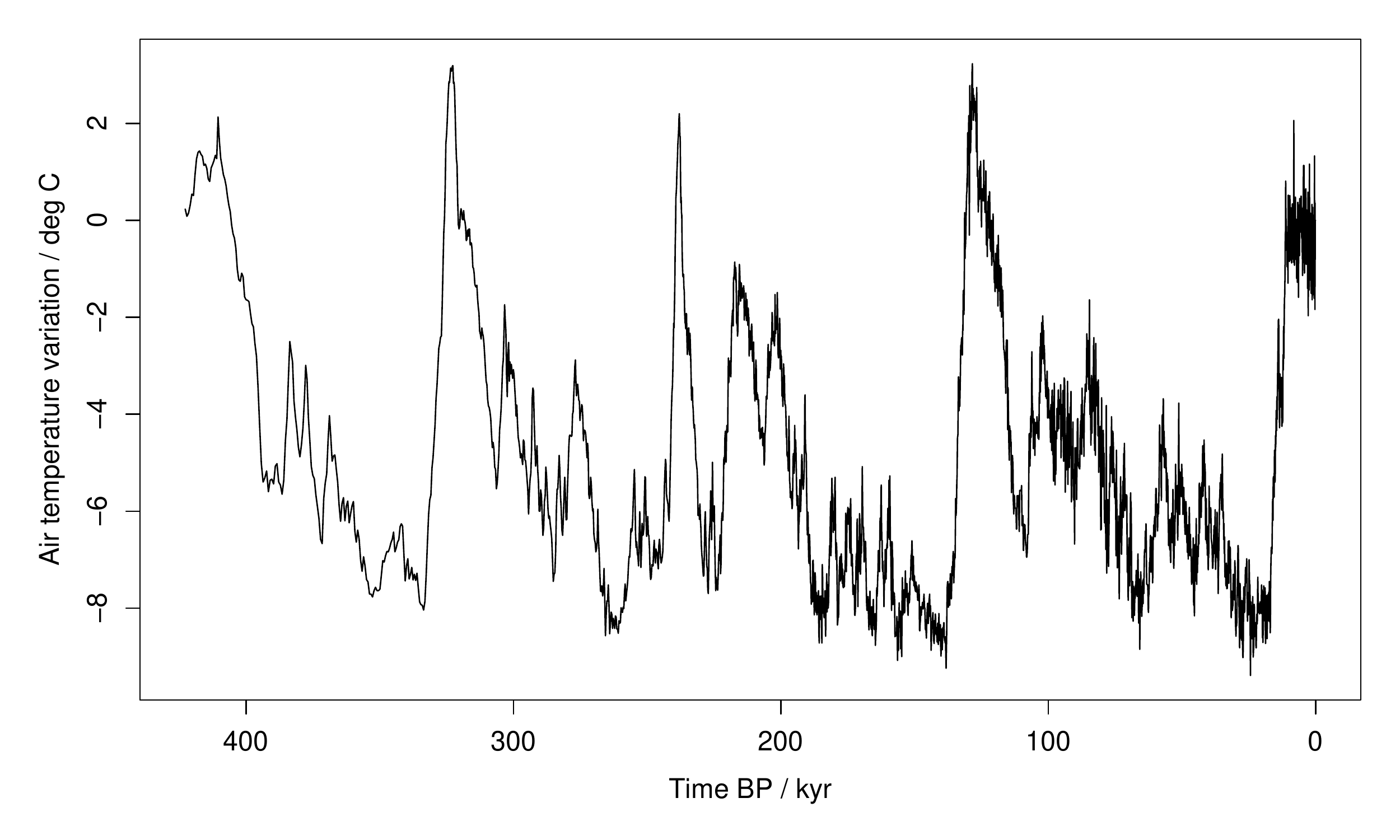}
\caption{Air temperature as measured in the Vostok ice core over the past 423\,kyr from data in Petit et al.~\citep{petit_etal_1999}.} 
\label{vostok_1999}
\end{center}
\end{figure}

\begin{figure}[htb]
\begin{center}
\includegraphics[scale=0.85, angle=0]{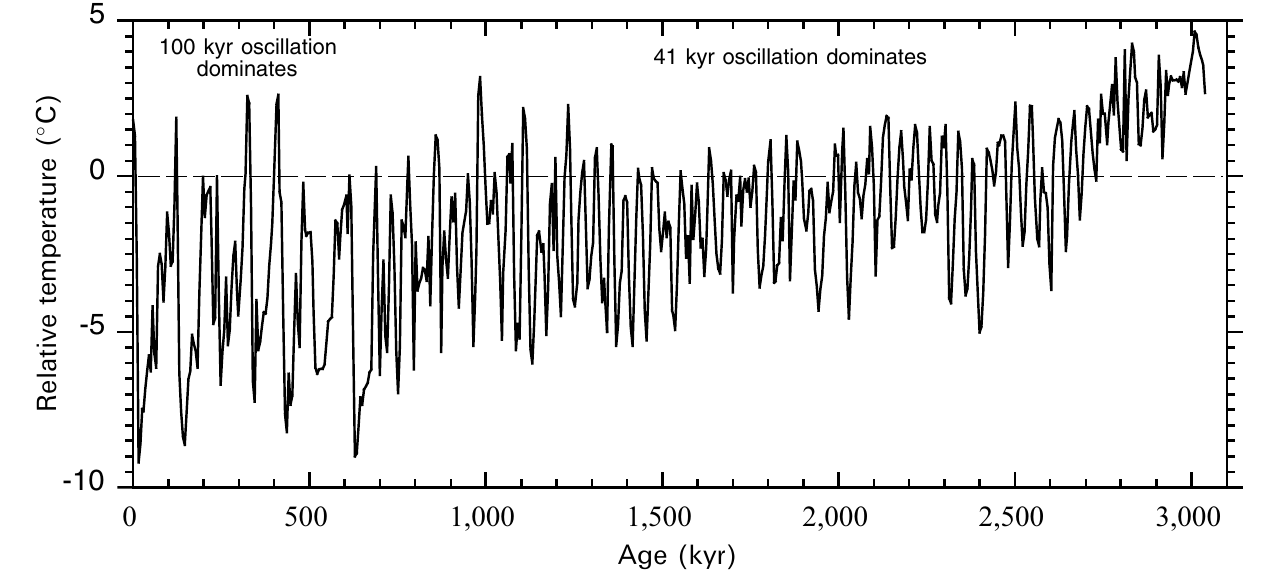}
\caption{Earth temperature variation over the past 3\,Myr. Note that time decreases from left to right (the present is at t=0).  Reproduced from Figure~1.6 from Muller \& MacDonald~\citep{muller_macdonald_2000} with kind permission of R.A\ Muller.} 
\label{muller_macdonald_2000_fig1_6}
\end{center}
\end{figure}

Atmospheric temperature has been recorded in Antarctic and Greenland ice cores going back as far as 800\,kyr BP at high temporal resolution (50--600\,yr per interval). Figure~\ref{vostok_1999} shows the temperature variation over the past 420\,kyr extracted from the Vostok ice core (Petit et al.~\citealp{petit_etal_1999}).  We clearly see cyclic behaviour on a timescale of about 100\,kyr. Figure~\ref{muller_macdonald_2000_fig1_6} shows the temperature extending back even further, using the \oproxy\ proxy measured in foraminifera.  The 100\,kyr oscillation period is still seen, but extends only back to about 700\,kyr BP. Before that a shorter period clearly dominates; a periodogram analysis identifies it as 41\,kyr, with lower power than the 100\,kyr one (Muller \& MacDonald~\citealp{muller_macdonald_2000}).  The relative power of peaks in the periodogram obviously depends on the time window selected.

The consensus is that these periodic variations reflect global temperatures and are caused by variations in the solar irradiance (insolation) at the Earth. (Although these variations are obviously both positive and negative, they are often referred to as causing ice ages.) The question is then what causes the variation in the solar irradiance.  A priori plausible astronomical mechanisms on timescales of a million years or less are intrinsic solar variability and variations in the eccentricity or inclination of the Earth's orbit about the Sun or in the orientation of the Earth's spin axis.  These will be discussed in sections~\ref{sun} and~\ref{orbit}.  Spiral arm and Galactic plane crossings as well as encounters with molecular clouds occur on much longer timescales, so can be ruled out a priori.
There is limited data on pre-quaternary (before 2.6\,Myr BP) climate, based for example on carbon and boron isotope ratios as proxies for \co2, and \oproxy\ and the Mg/Ca ratio in foraminifera as proxies for temperature (IPCC~\citealp{ipcc_2007}, section 6.3). There are no obvious periods in these.

\subsection{Periodicities in biological variation over the Phanerozoic}\label{phan_periods}

Many studies performed since the early 1980s have searched for periodicity in biological extinction or diversity data.  These have frequently used the compendium of fossil marine animal genera compiled by the late J.J.~Sepkoski\footnote{\url{http://strata.geology.wisc.edu/jack/}}.

\subsubsection{The 26\,Myr period}\label{26myr}

\begin{figure}[htb]
\begin{center}
\includegraphics[scale=0.55, angle=0]{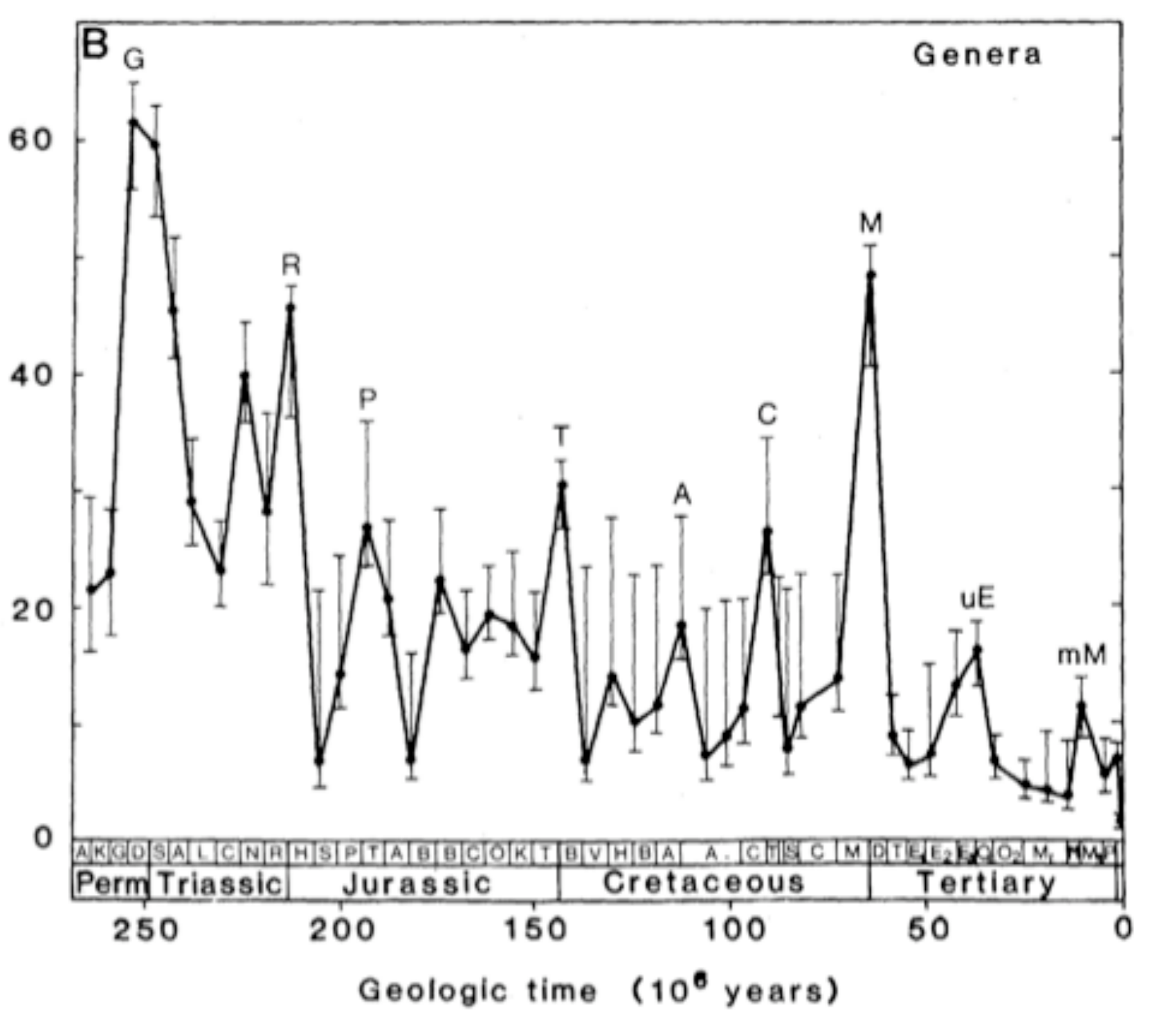}
\caption{Percentage of extinction among marine animal genera, reproduced from Raup \& Sepkoski~\citep{raup_sepkoski_1986} (Figure 1b). Reprinted with permission from the AAAS.} \label{raup_sepkoski_1986_fig1b}
\end{center}
\end{figure}

Raup \& Sepkoski~\citep{raup_sepkoski_1984} generated controversy in the mid 1980s when they claimed that there was a 26\,Myr period in the extinction record.  They used a database on 3500 marine animal taxonomic families spanning the 253--11\,Myr BP (Mesozoic and Cenozoic eras). The data are expressed as the fraction of extinctions in 39 stages, i.e.\ a time series with 39 points. When presented in this way, these data show 12 maxima (although one is extremely small) which appear to occur quasi-periodically.  These stages have a mean length of 6.2\,Myr, but because the exact durations are uncertain, they did not normalize the extinction fraction by the duration. This introduces uncertainties into the amplitudes of the time series.\footnote{The magnitude of these uncertainties is as large as the ratio of durations of the longest to shortest stages, a factor of about 2 judging from their Figure~1.}  They use the times of the 12 peaks as a set of delta functions as their raw data. Based on a Fourier and autocorrelation analysis they identify a significant ($p<0.01$) period between 27 and 35\,Myr. (The data are similar to those from their later study, reproduced here in Figure~\ref{raup_sepkoski_1986_fig1b}.)

Raup \& Sepkoski use a time domain technique for pulsed events to re-estimate the period. This technique has been used in other studies (e.g. Stothers~\citealp{stothers_1979}, Raup~\citealp{raup_1985a}) so I describe it briefly. The basic idea is to create a synthetic sequence of events at a certain period and phase and to calculate a goodness-of-fit which measures the standard deviation between 
each event in the real data and the closest one in the synthetic sequence.
This is repeated for a series of phases (e.g.\ $1,2,\ldots,P$) up to the full period ($P$), yielding the best-fitting phase and corresponding goodness-of-fit for that period. This is repeated for a range of periods and the best-fitting period/phase identified. This is a type of phase dispersion minimization technique, similar in principle to several others in the literature (e.g.\ Stellingwerf~\citealp{stellingwerf_1978}, Cincotta et al.~\citealp{cincotta_etal_1995}). 
A measure of significance can be defined using Monte Carlo simulations with random data:
from these we could calculate a probability distribution over the goodness-of-fit and so assign a $p$-value to the best-fit period/phase. Some applications instead ``normalize'' the goodness-of-fit based on the uniform distribution, using the procedure described by Stothers~\citep{stothers_1979}.  However, Lutz~\citep{lutz_1985} has shown that this overestimates the significance (see section~\ref{geomagnetic}).

One advantage of this technique is that because it identifies the nearest point in the synthetic series to each measured event, it is relatively insensitive to missing data. That is, it can identify a period in events even if some are missing. However, one must then provide an explanation for why some events are missing.

Using this technique, Raup \& Sepkoski~\citep{raup_sepkoski_1984} refine their 27--35\,Myr period
in the extinction data to 26\,Myr.
Although the authors acknowledge the significant uncertainty in both
the dating and the amplitudes of their data, they ultimately claim
this period to be significant. It gives rise to 10 cycles
over the 250\,Myr time span, such that the phased fit matches reasonably
well to only 8 of the 12 maxima (see Figure~1 of Raup \& Sepkoski~\citealp{raup_sepkoski_1984}).
Raup \& Sepkoski~\citep{raup_sepkoski_1986} revisited these data and
revised their conclusions to include just 8 significant extinction events,
but still with a 26\,Myr period (their Figure~1b is reproduced here as Figure~\ref{raup_sepkoski_1986_fig1b}.)
They report a high formal significance for this: $p<2 \times 10^{-4}$, although a visual inspection of the data shows that the extinction peaks are not evenly spaced. They reduce their significance to $p=0.05$ once dating errors are considered. It is important to realise that $p$ is just the probability of obtaining  
some data or statistic assuming that a particular null hypothesis (e.g.\ events drawn at random from a uniform distribution) is true.  It is {\em not} the probability that the null hypothesis is true, nor is $1-p$ the probability that some particular (and untested) alternative hypothesis (``periodicity'') is true. This will be discussed more in section~\ref{h_testing}.

The conclusion of Raup \& Sepkoski was criticized and strongly undermined by Stigler \& Wagner~\citep{stigler_wagner_1987} in a reanalysis of the data. Using the same time series analysis method, they agreed with the conclusion of Raup \& Sepkoski that the null hypothesis (that this time series is random) can be confidently rejected (from Monte Carlo tests they derive a $p$-value of $0.006 \pm 0.001$). However, rejection of this null hypothesis does not prove periodicity and Stigler \& Wagner demonstrate that non-periodic time series (such as a Moving Average), when analysed with the method of Raup \& Sepkoski, reveal significant periods. In other words, the time series method is not specific enough. (A similar conclusion was reached by Kitchell \& Pena~\citealp{kitchell_pena_1984}, discussed in section~\ref{nonperiodic}.)

It should be noted that the uncertainty in the dating is at least 6\,Myr, which is 23\% of the 26\,Myr period claimed by Raup \& Sepkoski. They reasonably argued that dating errors would blur out a real periodic signal, making it even harder to find, so that this cannot be used as an argument against the significance of a detection. But the situation is not quite so simple, because uncertainties are not necessarily random.  An important phenomenon is the ``Signor--Lipps'' effect: Because fossilization is rare, the last appearance of a fossil in a stratigraphic layer predates the final extinction, so a species/genus event may appear to go extinct in an earlier stage than it really did. (The effect is further enhanced by dating errors.)  Stigler \& Wagner showed that on account of the variable stage length in this data set, it is prone to showing an apparent period of around 26\,Myr when analysed with this time series analysis method. (The value of 26\,Myr is a consequence of the typical duration of the stages).
Raup \& Sepkoski~\citep{raup_sepkoski_1988} responded to this, but in their reply
Stigler \& Wagner~\citep{stigler_wagner_1988} stuck to and extended their original criticism.

It is worth asking 
what is the probability of detecting a period at all in the presence of dating errors?
This was examined by Heisler \& Tremaine~\citep{heisler_tremaine_1989}. They used a Monte Carlo technique to simulate time series with dating errors from a nominal time series with eight equally-spaced points, emulating the eight extinction events of Raup \& Sepkoski \citep{raup_sepkoski_1986}.  In each time series, points are randomly jittered by an amount chosen at random from a Gaussian distribution with zero mean and standard deviation $\Delta$.  Heisler \& Tremaine showed that if $\Delta$ is more than 13\% of the period (4.6\,Myr for a 26\,Myr period) then the period cannot be detected with a confidence above 0.90.  If $\Delta$ is increased to 23\% (6\,Myr, the average dating error) then the probability of detecting a period drops to 0.55. Heisler \& Tremaine therefore conclude that the period detected by Raup \& Sepkoski is either due to a statistical fluke or to a biasing of the data.\footnote{A biasing may come about (perhaps inadvertanetly) via ``cleaning'' of the data. The raw biological diversity data almost always have to be preprocessed, e.g.\ to remove single-occurence or poorly dated genera. Deciding what to remove based on which criteria is somewhat subjective, and the dependence -- robustness -- of any periods to these choices should be examined.} 
This conclusion seems unfounded, because a 50:50 chance of detecting the period is reasonable odds.
Interestingly, if the time errors are only 7\% of the period, then the period can still be detected with 100\% confidence (Figures 1 and 2 of Heisler \& Tremaine \citealp{heisler_tremaine_1989}). This might lead one to speculate that the dating uncertainties in the Raup \& Sepkoski data have been overestimated, thus reconciling the two analyses. But this is not the case, because the dating uncertainties are intrinsic to the placing of the extinctions in broad and variable-length (average of 30\,Myr long) stages. If the actual extinctions were not distributed at random within these, then this would be a case of a fluke which Heisler \& Tremaine refer to.

As I shall discuss in section~\ref{period_detect}, neither the Raup \& Sepkoski rejection of a null hypothesis of randomness, nor the Heisler \& Tremaine argument that a period could not be confidently detected, are sufficient to draw a conclusion about the existence of a period.  

A period of around 26\,Myr has been identified by other authors.  Rampino \& Caldeira~\citep{rampino_caldeira_1992} compiled data on seven different types of geological events including mass extinctions, orogenic events, sea-floor spreading and flood basalt volcanism.  The data are expressed as 80 delta functions spanning the past 260\,Myr. To create a contiguous data set more convenient for power spectral analysis they first smooth these data using a moving average (both Gaussian and top-hat functions are tried).  Power spectra are calculated from the Fourier Transform of the autocorrelation function using a Tukey window.  They identify several significant periods when analysing just the individual data sets, but a peak at 26\,Myr stands out when they combine all seven sets (as well as in some individual sets).  The significance is tested by seeing how many out of 1000 pseudo-random time series give a power equal to or greater than that detected at 26\,Myr.  They claim that these tests reject the null hypothesis at $p=1 \times 10^{-4}$, although how such a small $p$ can be achieved from just 1000 tests is unclear (as it would imply 0.1 tests showed an equal or higher power).  Moreover, the probability that the random data show a peak with this power or more at {\em any} period over 10--65\,Myr is 0.045, and this is surely the more relevant figure.

What is interesting about this study is that they combine data on different types of event. But are the data independent? After all, geological changes are used to define the boundaries in the geological timescale, which in turn is used to date the events. Rampino \& Caldeira nonetheless suggest that the periodicity is real and could reflect regularities in core/mantle dynamics (``pulsation tectonics''). Noting also that a similar period has been claimed for the impact cratering record, they suggest that the two may not be independent, i.e.\ impacts could cause geological change.
Napier~\citep{napier_1998} performed a similar study, combining data on impact craters, mass extinctions (from Raup \& Sepkoski \citealp{raup_sepkoski_1984}) and the geological events from Rampino \& Caldeira~\citep{rampino_caldeira_1992}.   He identifies a period of $27 \pm 1$\,Myr and links this to asteroid/comet impacts (or ``bombardment episodes'') driven by Galactic plane passages.

\subsubsection{The 62\,Myr period}\label{62myr}

\begin{figure}[htb]
\begin{center}
\includegraphics[scale=0.85, angle=0]{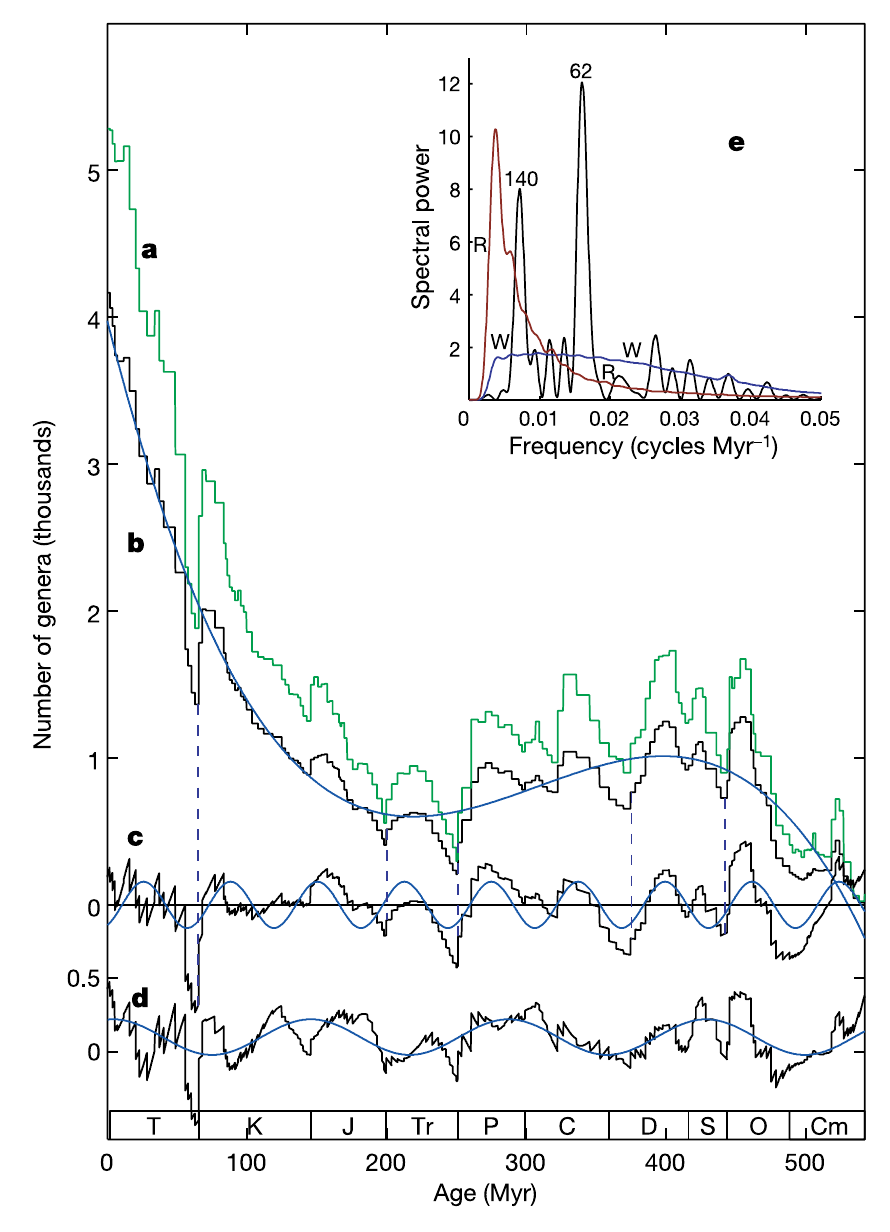}
\caption{Figure 1 from Rohde \& Muller~\citep{rohde_muller_2005}, showing the variation in genera diversity
before (green, a) and after (black, b) removing poor data. (c) shows (b) after removal of a third-order polynomial trend (with the 62\,Myr period sinusoid overplotted). (d) shows the same having also subtracted the 62\,Myr fit (with a 140\,Myr period sinusoid overplotted). (e) is the power spectrum showing two alternative backgrounds (W and R) for the significance calculation. Note that time decreases from left to right (the present is at t=0). 
Reprinted by permission from Macmillan Publishers Ltd.\ (Nature 2005).}
\label{rohde_muller_2005_fig1}
\end{center}
\end{figure}

Rohde \& Muller~\citep{rohde_muller_2005} examined the variation of the biological diversity in the Sepkoski~\citep{sepkoski_2002} database, recalibrated according to the timescale published by the International Commission on Stratigraphy (ICS 2004; Gradstein et al.~\citealp{ics_2005}). The data -- both the complete data set and one pruned of single-occurrence genera and ``poorly dated'' genera -- suggest a periodicity which is visually stronger after the removal of a slowly varying component with a third-order polynomial (see Figure~\ref{rohde_muller_2005_fig1}).  A power spectral analysis reveals a strong peak at $62 \pm 3$\,Myr and a weaker peak at $140 \pm 15$\,Myr.

When one talks about the ``significance'' of the detection of an event, such as a peak in a power spectrum, one is normally concerned with the probability that the event could have been caused by a ``random'' process. This is typically the noise associated with some background (what we would measure in the absence of a signal). But often the noise process is not well understood, so there is no unique definition of ``random'': we could draw events at random from a uniform or a Gaussian distribution, for example.  In classical hypothesis testing, one tries to reject the null hypothesis that the signal was produced by some specified background. 

In the case of power spectral analysis it is usual to calculate the significance of peaks in a power spectrum by comparing it to the power spectrum of the background process.  Rohde \& Muller use two different backgrounds: a complete randomization of the detrended data (R); and a randomization in consecutive stages (of length 27\,Myr), which preserves the short-term correlations (W); see Figure~\ref{rohde_muller_2005_fig1}. W reflects the fact that if the data are periodic then nearby points are not independent.  Both are examples of ``red noise'', in which the background is higher at lower frequencies. W of course has more power than R at high frequencies (and less at low), with the choice of stage length to use for W affecting its power spectrum (in the limit of small stages it will look like the R power spectrum; this illustrates that there is no simple definition of ``random background'' as a null hypothesis.)  The probability of these achieving a power as high as that seen in the 62\,Myr period {\em anywhere in the spectrum} is quoted as $<0.0013$ for R and $0.01$ for W. These decrease to around $1 \times 10^{-4}$ when we only consider the probability of getting a peak at 62\,Myr; that is, if we suspect a peak at this period (before we look at the data) and examine the evidence for a peak only at this period. This might be valid if one has independent evidence for a peak at this period and if one ignores any other peaks. But there is no prior reason to suspect a peak at 62\,Myr, so the higher -- yet still significant -- $p$-values should be used.  The 140\,Myr peak is only significant at $p<0.01$ if one limits the significance test to this period (otherwise $p=0.71$ and $0.13$ for R and W respectively), and then only for background model W. We should not regard this as a significant detection.

The plot of the fitted 62\,Myr sinusoidal curve to the data (Figure~\ref{rohde_muller_2005_fig1}) shows reasonable agreement in period and phase with the detrended data for much of the Phanerozoic, although the fit is quite poor over the past 150\,Myr. This is not inconsistent with finding a significant period, because even if this is a statistically significant detection, it does not automatically follow that it explains a significant part of the {\em total} variance in the data. All we have found is that it describes significantly more variance than the background process, and more than any other single period. The analysis has not ruled out that some other process may explain as much or even more of the variance in the data. Indeed, we see in Figure~\ref{rohde_muller_2005_fig1}b that the variation due to the long-term trend swamps the periodic variations, and may reflect a far more important aspect of biodiversity change.

Rohde \& Muller stress that the 62\,Myr period is only found when using the ICS 2004 timescale. They also ask whether this is a period in the {\em true} diversity or just in the {\em observed} diversity.  One subsequent study (Smith \& McGowan~\citealp{smith_mcgowan_2005}) has suggested that much of the signal is a selection effect (see section~\ref{reliability}).  
Incidentally, Rohde \& Muller were not the first to suggest a 62\,Myr period. Thirty years earlier, Thomson~\citep{thomson_1976} suggested there was such a period in the number of genera in various groups of lower vertebrates and certain invertebrates, although this very short article presents just a superficial analysis.

Methodologically the Rohde \& Muller result has been criticized by Omerbashich~\citep{omerbashich_2006} on three accounts: (1) interpolation of the time series to give uniform spacing (increasing the original 167 data points to 2170 points; this is necessary for using the Fast Fourier Transform, FFT); (2) the polynomial detrending prior to taking the power spectrum; (3) the use of zero padding in the Fourier Transform.  Interpolation can cause problems, e.g.\ by introducing short scale correlations in the data, but is acceptable as long as one realises that it corresponds to a smoothing at high frequencies.  Detrending is a standard procedure and seems valid, provided one recognises that it assumes that whatever causes the long-term trend is independent of whatever causes the periodic variations.  Zero-padding the time series to lengthen the time span is a common technique with the FFT to reach the maximum possible resolution of the power spectrum. Note that this does not (cannot) add information, so one should estimate the maximum resolution supported by the data and take care not to overinterpret the results.

Omerbashich reanalysed the original 167 points from Rohde \& Muller using the Lomb--Scargle (LS) periodogram (Lomb~\citealp{lomb_1976}, Scargle~\citealp{scargle_1982}) and without applying any detrending.  (LS works with unevenly sampled data and does not need zero padding.  It goes under various names, including Gauss--Van\'{\i}\v{c}ek spectral analysis.)
He found the 62\,Myr period to be insignificant at the $p=0.01$ level, and instead reports significant periods at 194, 140 and (according to his Figure~2a) 77\,Myr. However, this difference appears to be the result of not detrending, rather than the data preprocessing and spectral analysis technique used: Cornette~\citep{cornette_2007} found that a LS analysis of the original 167 data points, detrended, still gives rise to a significant period at 62\,Myr. In contrast, Cornette found no period to be significant via either LS or FFT if the data are not detrended.
This suggests that the detrending increases the sensitivity to the periods reported.
Cornette also concludes that the additional 140\,Myr period reported by Rohde \& Muller is not statistically significant. 
(This is not surprising, because this period is a quarter of the total time window, and thus sensitive to the exact detrending applied.)

Lieberman \& Melott~\citep{lieberman_melott_2007} also reanalysed the Rohde \& Muller biodiversity data and also detect the 62\,Myr period as significant (see also Lieberman \& Melott~\citealp{lieberman_melott_2009}).  They further analyse selected time windows of the biodiversity data: while the 62\,Myr periodicity is strong over the period 520--150\,Myr BP, there is no evidence for a significant period over the last 150\,Myr, which, as mentioned above, is not surprising when one inspects the time series fit to the data (Figure~\ref{rohde_muller_2005_fig1}). Thus any astronomical mechanism invoked to explain the 62\,Myr period must also explain why it did not significantly influence biodiversity in the past 150\,Myr. This could imply there are multiple mechanisms at work.  Lieberman \& Melott~\citep{lieberman_melott_2007} also examine extinction and origination data and find a significant period at 27\,Myr, although they caution that this might be an artefact of the data.
Melott \& Bambach (in preparation) also find a 62\,Myr period in other geological and biodiversity measures (private communication).

Many of the published articles on periods (26\,Myr or 62\,Myr) in biological diversity are based on the same database or updated versions thereof by Sepkoski. Melott~\citep{melott_2008} therefore looked for periods in the Paleobiology Database, a more recent compilation from many sources where particular attention has been made to correct for the preservation bias.\footnote{\url{http://paleodb.org/}} 
He finds a period of 63\,Myr (using correlation analysis, FFT and LS), consistent with the $62 \pm 3$\,Myr period found by Rohde \& Muller~\citep{rohde_muller_2005}, further checking that the two had the same phase (to within 1.6\,Myr, i.e.\ within the measurement uncertainty). Two other statistically significant peaks are disregarded as artefacts of detrending and variations in the interval length. There is no figure showing how the fit corresponds to the original data, so it is unclear whether there is a good fit over the whole Phanerozoic. 
Alroy~\citep{alroy_2008} looked for periodicity on data selected from the same database. He finds no significant autocorrelation and says that the power spectrum (Figure S2 in the online supplement) is consistent with white noise: There is no significant power around 26\,Myr, although he suggests weak evidence for a significant period ``somewhat longer than 62\,Myr''. (The definition of ``significance'' depends on a 95\% confidence band around the median power, the origin of which is not clear.)

In summary, there have been claims of a periods in the fossil record of 26\,Myr spanning 250--10\,Myr and of 62\,Myr spanning 520--150\,Myr (formal uncertainties in the periods are 1--3\,Myr). The former (and older) claim has been heavily criticised on grounds of data and methodology and I believe now has little credibility (more on this in section~\ref{ts_fun}). The 62\,Myr period is on a stronger footing, although less time has passed for this to be reanalyzed/criticized and many of the issues I'll discuss in sections~\ref{reliability} and~\ref{ts_fun} apply.

%%%%%%%%%%%%%%%%%%%%%%%%%%%%%%%%%%%%%%%%%%
\subsection{Periodicities in the geological record over the Phanerozoic}\label{geoperiods}

\subsubsection{Impact cratering}\label{craters}

Following the dramatic claim by Alvarez et al.~\citep{alvarez_etal_1980} that a meteorite or comet impact wiped out the dinosaurs at the K-T boundary 65\,Myr ago, there were many studies in the 1980s which looked for periodicity in the cratering record.  Rampino \& Stothers~\citep{rampino_stothers_1984a} claimed that the dates of 41 craters dated from 250--1\,Myr BP showed a dominant period at $31\pm1$\,Myr.  They also claimed that nine mass extinction events occurred at around the same times and that these could be coincident with Galactic plane crossings. (However, their Table 1 hardly supports this, with phase differences between extinctions and plane crossings ranging between $-14$ and $+11$\,Myr, i.e.\ essentially the whole period.)  Napier~\citep{napier_1998} identified a period of $12.5$\,Myr for 31 well-dated craters over the last 150\,Myr (although he suggests this may be a harmonic of a true period of 25--27\,Myr). Matsumoto \& Kubotani~\citep{matsumoto_kubotani_1996} suggest that there is period of 30\,Myr in the cratering record over the past 300\,Myr. After smoothing this to a continuous time series (in part to accommodate the dating errors), they claim it agrees in period and phase with the extinction data of Raup \& Sepkoski (\citealp{raup_sepkoski_1984}) to moderate confidence ($p=$\,0.03 to 0.07).

Several other studies have rebuffed such claims.  For example, Grieve et al.~\citep{grieve_etal_1985} showed that several different periods could be fit to the cratering record, depending on the events included and the date scale used. Given the dating errors and incompleteness of the record, there are many data preprocessing decisions to be taken and so a lot of inherent flexibility in the models which can be fit: Even if an analysis on a single set of data gives statistically significant results, one should question how robust this is to small perturbations of the data.

Some criticisms have been more direct.  Stigler~\citep{stigler_1985} refers to the statistical argument of Rampino \& Stothers~\citep{rampino_stothers_1984a} -- where they claim a significant correlation ($\rho=0.996$) between the dates of nine extinction events and Galactic plane crossings -- as ``seriously misleading''. One reason is that the correlation coefficient of any two monotonically increasing series is bound to be high. For example, the correlation between the dates in Myr of the nine extinction events ($11, 37, 66, 91, 144, 176, 193, 217, 245$) and the first nine prime numbers ($2, 3, 5, 7, 11, 13, 17, 19, 23$) is 0.986, even though they are entirely unrelated. Although the reply by Rampino \& Stothers (\citealp{rampino_stothers_1984b}) notes that only 0.4\% of random data achieves $\rho > 0.996$, a difference in $\rho$ of just $0.01$ is a very unstable basis on which to build an astronomical theory of mass extinctions (even if one neglects the large dating errors and the huge uncertainties in the Galactic plane crossing model).

Grieve et al.~\citep{grieve_etal_1988} examined how reliably periods could be recovered from simulated time series data in the presence of dating errors and superimposed random events. Given what we know about the orbits of minor bodies in the solar system, it is highly unlikely that {\em all} large impacts have a single external trigger. For example, impacts would not {\em only} occur -- nor would be guaranteed to occur -- when the Sun encounters a molecular cloud. Therefore, even if there is a periodic component in the impact distribution, it would be superimposed on a random one. Grieve et al.\ show that if a time series is a 50:50 mix of periodic and random events, a period can only be detected at a 99\% confidence level 
if the dating errors are less than 10\% of the period.  Given the magnitude of dating errors in the cratering record ($\sim 5$\,Myr), this implies that
true periods shorter than 50\,Myr are not detectable with {\em high} confidence. Like Heisler \& Tremaine~\citep{heisler_tremaine_1989} after them (see section~\ref{phan_periods}), they therefore conclude that many periods claimed in the literature are statistically fortuitous.
Using a time-domain method to compare data sets with simulated sequences of delta functions at different periods, Grieve et al.\ conclude that there is no strong evidence for a periodicity of 30\,Myr in the cratering record over the past 220\,Myr. Although they find periods in the vicinity of 16 and 20\,Myr (depending on the data set used), they indicate that these are not real due to both age uncertainties, as well as the siderophile composition of several impactors indicating that they were not comets (so should not have been included in the analysis). 

Jetsu \& Pelt~\citep{jetsu_pelt_2000} demonstrated that the ``human signal'' of rounding ages (e.g. 66.7 to $67 \pm 1$) can, contrary to what we might expect, produce a spurious periodic signal. They examine in particular the impact cratering record and conclude that this rounding is responsible for the period of 28.4\,Myr found by Alvarez \& Muller~\citep{alvarez_muller_1984}. Using Monte Carlo significance tests they find no reliable period in any of six different cratering data sets using a variety of period search methods.

\subsubsection{Geomagnetic reversals}\label{geomagnetic}

The orientation of the geomagnetic field is preserved in some rocks. From this record it has been found that the field has undergone frequent reversals.  The frequency of reversals has varied considerably, from several per Myr in the recent past to long durations of no reversal (e.g.\ the Cretaceous Superchron lasting from 82--118\,Myr BP). 
It is believed that these reversals are a result of intrinsic instabilities in the Earth's dynamo, although Muller \& Morris~\citep{muller_morris_1986} proposed a mechanism by which an extraterrestrial impact could flip the field.
Some authors have claimed the reversals to be periodic. Negi \& Tiwari~\citep{negi_tiwari_1983} report evidence for several periodicities in these reversals, the most significant and longest being 285\,Myr.
Raup~\citep{raup_1985a} -- using the phase dispersion minimization technique with the normalization of Stothers~\citep{stothers_1979} described in section~\ref{phan_periods} -- claimed that there was a 30\,Myr periodicity in the rate of reversals over the past 165\,Myr (when binned into blocks of 5\,Myr duration). However, Lutz~\citep{lutz_1985} showed that this was an artefact of the method, the specific period being determined by the record length. Raup subsequently retracted his result (Raup~\citealp{raup_1985b}), although Stothers~\citep{stothers_1986} continued the claim that the 30\,Myr period was significant. 

In summary of this section, several periods (e.g.\ 12.5, 28.4, 30, $31 \pm 1$\,Myr) have been claimed in one or more geological records. (Other papers not discussed here come up with yet other periods over this range.) Some agree within the uncertainties or at least are close to the 26\,Myr period in extinctions from Raup \& Sepkoski~\citep{raup_sepkoski_1984}, \citep{raup_sepkoski_1986}, leading many to suggest a causal connection. As with the periods in the biological records, many of these studies have been criticized on methodological grounds and on account of uncertainties. Moreover, several studies find no evidence for periods. In the next section (\ref{nonperiodic}) I will look beyond simple periodic explanations, and in the section after (\ref{reliability}) I'll examine some of the data uncertainties.

%%%%%%%%%%%%%%%%%%%%%%%%%%%%%%%%%%%%%%%%%%
\subsection{Complex systems: non-periodicity and multiple mechanisms}\label{nonperiodic}

Although there is a suggestion of periodicity in the biological and geological records, a single period is often not a good fit, at least not over long time scales.  We saw this already for climate data over the past 3\,Myr (Figure~\ref{muller_macdonald_2000_fig1_6}), where there is a clear change in the period. Attempting to derive a single period persistent over the whole time series may give a meaningless result, depending on the method used. Rampino~\citep{rampino_1998} has suggested that a better periodic fit to extinction and cratering data is obtained by allowing a variable period. Alternatively one could allow phase shifts with a constant period.  Of course, both of these would require a causal mechanism. In section~\ref{solar_motion} we will see that the orbit of the Sun about the Galaxy naturally accommodates a variable period of its vertical oscillations through the Galactic disk, either due to spiral arms or the $R$-dependence of the potential.

There are many studies in the literature which conclude that biodiversity does not vary periodically.
In his review, Bambach~\citep{bambach_2006} argues that in the Sepkoski genus-level database there are only three events which have a level of extinction significantly higher than the background, thus challenging the claim that mass extinctions are common enough to even attempt to describe them as periodic. 
(Some have made claim of a ``big five'' of mass extinctions, other just a single ``big one''. This depends on how we define ``big'' or ``significant''.)
Based on the diverse impact of lower intensity extinctions on biological diversity, he concludes that these are unlikely to have a single cause. This is an important point, because if several mechanisms are at work, then even if one or more of these are periodic, attempts to derive a single period will give spurious results.
Bambach also concludes that the varability and the peaks in the extinction rate are not a result of incompleteness in the fossil record.

Many authors have argued for the existence of multiple extinction mechanisms.
White \& Saunders~\citep{white_saunders_2005} note that large impacts and mass volcanism have occurred much more frequently in the past 300\,Myr than have mass extinctions, so neither of these {\em in isolation} could have caused the mass extinctions. They present a statistical analysis which shows that a few random coincidences of these two mechanisms is not unlikely, and could explain the frequency of mass extinctions.  In support of this, Hallam~\citep{hallam_2004} notes that there are many large impact craters for which no contemporary mass extinction has been identified. 
Arens \& West~\citep{arens_west_2008} likewise argue that a simultaneity of volcanism and impacts (or, more generally, what they call ``press'' and ``pulse'' events) is necessary to explain mass extinctions.

It is well known that apparent periodicities in data can be produced by non-periodic processes. Chaotic phenomena, for example, can exhibit quasi-periodic signals.  It is not even obvious that extraordinary events such as mass extinctions require an external driver (whether extraterrestrial or not, or periodic or not).  
From an analysis of the correlation between biodiversity and extinction, in particular the observation that a high rate of extinction tends to follow times of large biodiversity, Alroy~\citep{alroy_2008} concludes that at least some variation in the fossil record can be explained by purely
ecological interactions (e.g.\ predation and competition). However, this diversity--extinction correlation is weak ($\rho=0.44$).
Numerical models have also shown how, in such interdependent systems, small perturbations can produce large extinctions and large perturbations may result only in small extinctions (e.g.\ Plotnick \& McKinney~\citealp{plotnick_mckinney_1993}). The implication is that we should not necessarily look for a linear correlation between cause and effect.  It has further been suggested that evolution may be a self-organized critical phenomenon arising from the interaction between species (e.g.\ Kauffman \& Johnsen~\citealp{kauffman_johnsen_1991}). Such dynamical systems can show highly nonlinear responses, with extreme events an almost inevitable consequence of the system dynamics. Evidence supporting this idea is the power-law distribution of the magnitude of extinction events and taxa lifetimes (Newman~\citealp{newman_1997}), although the data are noisy and other fits are possible.  Stanley~\citep{stanley_1990} has suggested that apparent periodicities in biodiversity are a result of delayed ecosystem recovery from mass extinction and thus a feature of the system rather than any driver.

Kitchell \& Pena~\citep{kitchell_pena_1984} fit various time-domain models to the extinction data of Raup \& Sepkoski (\citealp{raup_sepkoski_1984}) and found that a stochastic autoregressive (AR) model provides a good fit (just as Stigler \& Wagner~\citealp{stigler_wagner_1987} found that a moving average process is a good model for these data; section~\ref{26myr}).  In this AR model, each value of the time series is a linear combination of the previous five events. The system retains a memory and can display pseudo-periodic behaviour, even though there is no external driving force. Kitchell \& Pena find that AR model fits the data better than either a continuous periodic variation or a model with periodic impulses. Indeed, they claim that the lack of strict periodicity and the large variations in extinction magnitude rule out a deterministic explanation of the data, which in turn suggests that biodiversity is a dynamic phenomenon.  Of course, just because the AR model fits the data well, this does not ``rule out'' the periodic model -- we must compare the posterior probabilities of the models (section~\ref{h_testing}) -- and does not rule out the presence of an external driver.  Yet the onus is on us to show that the periodic model is considerably more likely, and the good AR fit argues against a single external driver being the sole or even dominant cause.

\subsection{How reliable are the data, how appropriate are the methods?}\label{reliability}

Geological data are far from perfect. One of the most significant issues is that of dating. There exist different dating scales and one of the more robust results reported -- the 62\,Myr periodicity of Rohde \& Muller~\citep{rohde_muller_2005} (section~\ref{62myr}) -- was only found when the Sepkoski~\citep{sepkoski_2002} data were recalibrated on the ICS 2004 timescale. There may be good reasons to assume that this is the most accurate timescale to date, but that was presumably thought of earlier timescales too. In addition to such calibration uncertainties, there are significant random errors in dating extinctions or impact craters, which can extend to tens of Myr. 

Beyond dating, there is the problem of the incompleteness of the geological record.  
Concerning the impact record, presumably not all craters have been found, some have eroded or been subducted into the Earth's interior, and the amount of land surface available for impacts has varied over time.
Similarly, the efficiency of fossilization depends on environment and species, thus giving rise to biases.  Fossilization is anyway rare, and as I already discussed in section~\ref{26myr} it gives rise to the Signor--Lipps effect. In addition to possibly producing spurious periods under some circumstances (as already discussed), this effect tends to smooth peaks in the extinction record, making extinctions appear more gradual than they really were (e.g.\ Hallam~\citealp{hallam_2004}, chapter 3).  
Fossils have also not been searched for uniformly over the globe, introducing a geographical bias (compounded by the fact that
the continental plates have migrated with time).

One must also ask how reliably the recorded variable measures the phenomenon of interest.  By measuring the size of an extinction through the number of genera which survive, we put genera in which almost all species died on an equal footing with genera which were unaffected. A different measure will be obtained if families or the total biodiversity are used instead (section~\ref{geobiodata}).  All of this is measured via fossils, but to what extent do the data represent variations in biological life rather than variations in their preservation in the fossil record?  If we are open to the idea of a mechanism which has a widespread impact on life, then we should be equally open to the idea of a mechanism which can affect preservation.  Peters \& Foote~\citep{peters_foote_2002} showed that much of the observed variability in the marine fossil record can be explained by variations in the amount of rock exposed for fossil preservation, rather than variations in the extinction rate itself (see also Smith~\citealp{smith_2007}).  Smith \& McGowan~\citep{smith_mcgowan_2005} give evidence suggesting that this is the source of the biodiversity variation reported by Rohde \& Muller~\citep{rohde_muller_2005} (see section~\ref{62myr}).  They note that long-lived taxa, which are less affected by this sampling bias, do not show periodic variations.
Rock exposure variations could be due to changes in sea level, themselves plausibly driven by plate tectonics or climate change (locking ice into glaciers and ice caps).
Of course, one must still explain why short-lived taxa and/or sea levels appear to vary periodically and whether preservation issues
interfere with periods on all timescales.

Our variable of interest is often measured indirectly via a proxy, which must be calibrated. With ice cores, for example, one measures \oproxy\ as a function of ice depth, yet \oproxy\ must be converted to temperature (using calibration terms which depend on geographical location) and depth must be converted to time.  Our proxies/samples may also be contaminated.  For example, Royer et al.~\citep{royer_etal_2004} showed that the \oproxy\ measured in marine carbonates has to be corrected for water pH.  Patterson \& Smith~\citep{patterson_smith_1987} claimed that 75\% of the extinctions in families reported by Raup \& Sepkoski~\citep{raup_sepkoski_1984},~\citep{raup_sepkoski_1986} (section~\ref{26myr}) are artifical extinctions introduced by taxonomic definitions.
This includes ``pseudo-extinctions'' of species which do not die out but rather evolve into something else.

In terms of the techniques, there are many ways of analysing a time series.  Many appear appropriate, but they can produce different results.  There are decisions of how to calculate the significance, what to accept as significant (and so report anything at all), what filtering to use, what data to omit, etc. 
Several authors (e.g.\ Grieve et al.~\citealp{grieve_etal_1988}, Napier~\citealp{napier_1998}, Lieberman \& Melott~\citealp{lieberman_melott_2007}, Melott~\citealp{melott_2008}) have found statistically significant periods in data which they then argue away (perhaps quite reasonably) on other grounds.

Is our choice of data appropriate? If extinction data are essentially a contiguous time series, does it make sense to focus on selected large extinction events and try to explain just these? This would rule out testing mechanisms which are capable of explaining both low and high amplitude events.  
For example, Alroy~\citep{alroy_2008} find that the distribution of extinction and origination rates of marine invertebrates over the whole Phanerozoic can be fit well with a log normal (after detrending). This doesn't necessarily imply a single common mechanism.
But do we need to invoke occasional catastrophic triggers if in fact large amplitude events can be explained as the tail of a more mundane process?

This summary of the issues is not intended as blanket scepticism, but rather as a reminder that there are choices to be made in data analysis. It is not as objective as it sometimes appears to be.  Time series analysis is a complicated business: there are decisions of which data to include, what preprocessing to do, which methods to use and which significance tests to apply.  We should question the assumptions and identify the uncertainties in order to examine what impact they have on potentially far-reaching conclusions.

\subsection{How independent are the studies?}

Numerous articles have identified supposedly significant periodicities with a period in the range 25--33\,Myr.  This has led many authors to speculate an astrophysical cause, partly because of the lack of a plausible terrestrial mechanism for such a periodicity. Despite various criticisms, the sheer number of studies converging on a similar values is noteworthy.  But are they independent? First, many studies use the same data sets and dating system, so are subject to the same systematic errors and sampling biases. Second, some studies use the same methods, some of which have been demonstrated to have deficiencies. Third, what is taken as ``evidence'' or ``significance'' is often inadequate. Fourth, many other studies find different periods or no periods at all.

A phenomenon worth noting is the ``band wagon'' effect, in which the presence of a published value biases authors' analyses (consciously or otherwise) and the conclusions they choose to publish towards confirming that published value. Studies with results lying far from the current trend may not be published at all.  In the current context, the presence of a mechanism which has a period near to one which the data could support may play a similar role.  
The band wagon effect is described in the context of distance measures to the Large Magellanic Cloud by Schaefer~\citep{schaefer_2008}, who notes that the estimates are far more consistent with each other than expected based on the reported uncertainties in the individual estimates.

%%%%%%%%%%%%%%%%%%%%%%%%%%%%%%%%%%%%%%%%%%%%%%%%%%%%%%%%%%%%%%%
\section{Issues in time series analysis}\label{ts_fun}

\subsection{Hypothesis testing}\label{h_testing}

The studies discussed above are concerned with assessing evidence for periodicity in time series data. This is an example of hypothesis testing. Probably the most common approach to this now in use is that developed by R.A.\ Fisher~\citep{fisher_1925}. The general idea is to define a null hypothesis, a model for producing the data in the {\em absence} of the effect one is investigating. We then calculate the probability that the data (or rather a statistic based on them) are predicted by this model. If this probability is low, we ``reject'' the null hypothesis at this confidence level, which suggests that some alternative hypothesis may be more likely to explain the data. As a simple example, imagine we have two groups of people, one which has received special training, the other not, and we
wish to assess the impact of the training on scores in an exam.
The typical null hypothesis states that the training makes no difference. If we assume the scores of the individuals in each group are distributed according to a Gaussian,  we would perform a $t$-test to examine whether the means of the two groups differ by a significant amount, where ``significant'' is referred to the pooled standard deviation of the two groups. The result is the ``$p$-value'' or $P(D|H_0)$, the probability of observing the data we did, $D$, given that the null hypothesis, $H_0$, is true. (This is sometimes called the ``likelihood'' of the hypothesis.)

In this simple example the null hypothesis ($H_0$; means equal) is the complement of an alternative ($H_A$; means unequal) which is of interest. The two hypotheses cover all possibilities, so the alternative hypothesis is implicit in the definition of the null. This is not generally the case, however. In the case of assessing the significance of peaks in a power spectrum of a (zero mean) time series (e.g.\ the biodiversity data shown in Figure~\ref{rohde_muller_2005_fig1}), we could define a null hypothesis as a time series with the same sampling drawn from a Gaussian distribution with mean zero and standard deviation, $\sigma$, set equal to the standard deviation in the original data. We can then measure its power spectrum (or rather the probability distribution of the power at any period via Monte Carlo simulations) and calculate the $p$-value of this null hypothesis. If $p$ is small (typically we require $p < 0.01$) we ``reject'' the null hypothesis. Many people automatically assume that this therefore ``accepts'' the alternative hypothesis. But this is not the case, because the alternative here is not the complement of the null. Indeed, we haven't even specified (let alone tested) the alternative! All we've done is assign a low probability to {\em one} specific null hypothesis. There might be other null hypotheses which predict the data with a higher probability. There are certainly other ways to specify the null (a different value of $\sigma$, non-Gaussian noise, retain short-term correlations etc.).

Ideally we try and reject several different null hypotheses, as Rohde \& Muller did with their W and R background models.  But the model of interest (periodic variability) is not tested in orthodox hypothesis testing. The null hypothesis may be constructed using some properties of the measured data (e.g.\ the time sampling), but the data themselves are not tested.

Orthodox hypothesis testing has the further curiosity that it assesses the probability of getting data that are not observed. In the $t$-test example above, we don't actually calculate the probability of getting the data under the null hypothesis: the probability of observing a specific value from a continuous distribution is infinitesimally small. Instead, we calculate the probability that the means differ by the measured amount {\em or more}. Likewise with the power spectra: we would calculate (via Monte Carlo simulations) the probability that the null hypothesis can produce the observed power {\em or more}.  But why should we be interested in the probability of observing data we never actually saw?  This issue goes to the heart of criticisms of hypothesis testing (and the limitation of ``proof by contradiction'' or ``falsifiability''), which have been discussed extensively elsewhere (e.g.\ Berger~\citealp{berger_2003}, Jaynes~\citealp{jaynes_2003}, Christensen~\citealp{christensen_2005}, section 1.4 of Sober~\citealp{sober_2008}).

A yet more important point about hypothesis testing is the interpretation of the $p$-value.  If we calculate a low $p$-value for a null hypothesis $H_0$, then we have found that $P(D|H_0)$ is low. This does not mean that $P(H_A | D)$ is high! (Some articles nonetheless interpret $1-p$ as the probability that the alternative is true. This is wrong.)  Strictly we cannot even ``reject'' $H_0$; for this we would need to know $P(H_0|D)$ which is not the same as $P(D|H_0)$. We can illustrate this with a simple example. Imagine you draw a card at random from a deck of cards and it's the ace of spades. This is $D$. The probability of drawing this, assuming it's a normal deck ($H_0$), is $P(D|H_0) = 1/52$. But you are unlikely to tell me that the probability that the deck of cards is normal [$P(H_0|D)$] is $1/52$. Or take a more extreme example: a particular person wins the lottery with a chance of 1 in $10^8$. Would you claim that the probability that the lottery is fair is $10^{-8}$?  We have a similar problem with scientific data: the probability of getting the data we actually observe is very small under almost any hypothesis (vanishingly so with continuous variables).\footnote{This is why Fisher hypothesis testing is forced to calculate a $p$-value for a range of values,  e.g.\ probability of getting that power {\em or more}.} 
The point is that even a low $P(D|H_0)$ or $p$-value may provide more support for the null hypothesis than for any other alternative. We cannot know how low $p$ should be in order to reject the null hypothesis (lottery is fair) without knowing the $p$-value of alternative hypotheses (lottery is rigged, lottery is fair but someone bought all of the tickets etc.). If a data point lies 10$\sigma$ from the mean of a Gaussian, I can only say the data didn't come from that Gaussian if I accept there is an alternative origin.  How likely we think there to be an alternative is quantified using the {\em prior probability} of the model. 
This is the probability that the model is true independent of (before using) the specific data $D$ (Sivia~\citealp{sivia_1996}, Jaynes~\citealp{jaynes_2003}, Gregory~\citealp{gregory_2005}).\footnote{The prior is determined by previously obtained data and experience, which always influence our choices. As an example, the fact that we don't look for periods in impact cratering on periods less than a year, say, is equivalent to saying that our prior probability for such periods is zero.}  Only if the prior probability of the alternative is very small might this measurement give evidence in favour of the Gaussian origin.

The solution to this dilemma is to do direct model comparison, that is, to compare the likelihoods, $P(D|H_i)$, for two explicit models (their ratio is called the ``odds ratio''). If we give the two models equal prior probabilities, $P(H_i)$, then the one with the largest likelihood better predicts the data. Yet this still doesn't give the {\em posterior probability}, $P(H_M|D)$, for the hypothesis of interest ($H_M$, e.g.\ periodicity at some period). We can only calculate this if we know the complete set of alternative models, $H_i$. We can then use Bayes' theorem (which follows from the basic axioms of probability)
\begin{eqnarray}
P(H_M | D) &=& \frac{ P(D | H_M) P(H_M) }{ P(D) } \nonumber \\
                &=& \frac{ P(D | H_M) P(H_M) }{ \sum_i P(D | H_i) P(H_i) }
\label{eqn:phd}
\end{eqnarray}
The Bayesian approach to hypothesis testing is to explicitly test (calculate $P(D | H_i)$) all plausible models for the data, including the model of interest. Only in this way can we calculate the quantity we are actually interested in, $P(H_M | D)$. This avoids having to calculate the probability of observing data we might have seen but did not. Although this approach overcomes the limitations of orthodox hypothesis testing, it presents a new problem, namely the need to specify all plausible alternative models (all those which don't have a vanishingly small prior probability).  In most real-world problems it is almost impossible to define all plausible alternatives.  That is, the model space is incomplete. But at least this encourages us to define and test as many plausible alternatives as we can think of.\footnote{We also need to assign a prior probability to each model. This can be difficult in practice and is often criticized by orthodox statisticians. But it is arguably more honest than ignoring the alternatives altogether.}

\begin{figure}[htb]
\begin{center}
\includegraphics[scale=0.40, angle=0]{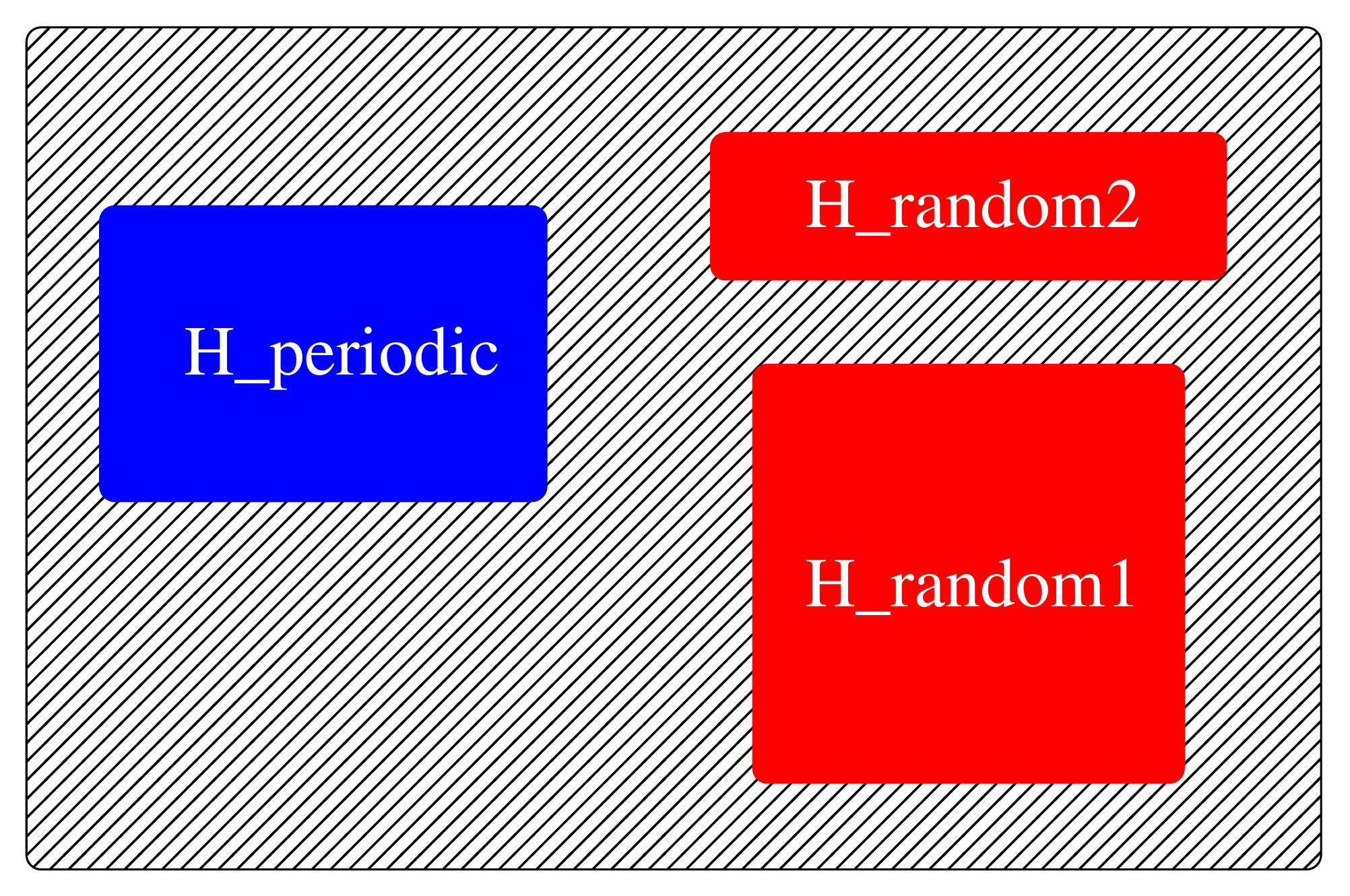}
\caption{Schematic illustration of hypothesis testing. The hatched black box shows the full hypothesis space. The blue box indicates the space covered by a certain periodic hypothesis and the two red boxes represent two alternative hypotheses for random models. In orthodox hypothesis testing for periodicity, one of the random model hypotheses will be considered unlikely if $P(D|H_{\rm random1})$ (the $p$-value) is low, even though this does not imply $P(H_{\rm random1} | D)$ is low. Moreover, this alone cannot imply that $P(D|H_{\rm periodic})$ is high, because generally there are other untested hypotheses such as H$_{\rm random2}$.}
\label{hypotest}
\end{center}
\end{figure}

The fundamental difference between orthodox hypothesis testing and the Bayesian approach is that the former tests the probability of a single null hypothesis whereas the latter always compares the probabilities of two or more alternative hypotheses. We can only ``reject'' a hypothesis if a better candidate is available: The unlikeliness of the data itself is not enough. This limitation of orthodox hypothesis testing I shall refer to as {\em incomplete inference}.  Rejecting a null hypothesis may be a useful first step, but only if it fills a large part of the hypothesis space (Figure~\ref{hypotest}).  If, in practice, we interpret a low $p$-value as evidence against the null, then it is because we implicitly assume the alternatives to have low prior probabilities.

This idea carries over to time series analysis. We can find the power spectrum of any data set (the Fourier Transform is just a basis function projection). But the peaks are only interesting if they cannot be produced by any other non-periodic model which is plausibly responsible for the data. 
There are methods of significance estimation for time series analysis which do not depend on $p$-values for the probability of unobserved data (e.g.\ Gregory \& Loredo~\citealp{gregory_loredo_1992}, Sturrock~\citealp{sturrock_2008}).  These typically result in lower significance estimates than do $p$-value estimates.

As an aside, there are of course many examples in science where doubt has been cast on a theory on the basis of a poor fit to data, without their being a concrete alternative defined at the time. Examples include the perihelion precession of Mercury (which couldn't be satisfactorily explained by Newtonian mechanics), or the ultraviolet spectra of black bodies (which were incorrectly predicted by classical physics). In essence, a low $p$ has sometimes been essential for motivating the search for alternative hypotheses, even though a formal approach to deciding how low $p$ should be before searching for alternatives (rather than repeating the experiment or reassessing the measurement errors) presumably has been rarely adopted. Nonetheless, a proper statistical approach is still required in order then to compare the alternatives.

\subsection{Reconciling a period detection with its non-detectability}\label{period_detect}

In section~\ref{26myr} I discussed the detection of a significant periodicity in extinction data by Raup \& Sepkoski~\citep{raup_sepkoski_1984}~\citep{raup_sepkoski_1986} plus the criticism by Heisler \& Tremaine~\citep{heisler_tremaine_1989} that, given the large dating errors, the probability of detecting a true period in these data is very small.  How can we reconcile these two claims? It turns out that the two articles are testing different hypotheses. Let $D$ represent the detection of a period in some time series data by some method.
Raup \& Sepkoski calculate the probability, $P(D|H_r)$, of detecting the period under the null hypothesis that the data were generated by some random process, $H_r$. Because this probability is low they reject $H_r$.  Heisler \& Tremaine, by examining the recovery rate of a period in simulated time series with timing errors, evaluate $P(D|H_p)$, where $H_p$ is the hypothesis that the process which produced the data is periodic. Because it's low they say one is unlikely to detect the period, hence the apparent period must be caused by something else (e.g.\ noise).  Yet neither study is sufficient to decide whether the data give evidence for $H_p$, because neither calculate $P(H_p|D)$.

$H_r$ is defined by the Monte Carlo method with which Raup \& Sepkoski generate random time series, from which they calculate their $p$-value.
They use the low value of $p$ to (1) reject $H_r$ and (2) infer that $H_p$ is true.  As described in the previous section, these two inferences make the additional assumption that $H_r$ is the only alternative to $H_p$. This is not true, because we could have defined other random processes to calculate the $p$-value. If we are nonetheless generous to Raup \& Sepkoski and assume that $H_r$ is the only alternative to $H_p$ then we can write $H_r = \overline{H_p}$ (the horizontal bar means ``not'').
The two hypotheses under consideration are then related by
\begin{equation}
P(D) = P(D | H_p) P(H_p) + P(D | \overline{H_p}) P(\overline{H_p}) 
\label{eqn:probsum}
\end{equation}
where $P(D)$ is the probability of detecting the period at all (under either hypothesis).
We are interested in 
$P(H_p | D) = 1 - P(\overline{H_p} | D)$. This is related
to the above quantities via Bayes' theorem
\begin{equation}
P(H_p | D) = \frac{P(D | H_p)P(H_p)}{P(D)}
\label{eqn:bayes}
\end{equation}
Substituting equation~\ref{eqn:probsum} into this gives
\begin{equation}
P(H_p | D) = \frac{P(D | H_p)P(H_p)}{ P(D | H_p) P(H_p) + P(D | \overline{H_p}) P(\overline{H_p})}
\label{eqn:bayes2}
\end{equation}
If we have no reason to prefer $H_p$ over $\overline{H_p}$, then we could set their prior probabilities to be equal.
They then cancel out of the above equation leaving
\begin{equation}
P(H_p | D) = \frac{1}{1 + \frac{P(D | \overline{H_p})}{P(D | H_p)}}
\label{eqn:posteqprior}
\end{equation}
Thus in order to decide whether the data, $D$, favour $H_p$ over $\overline{H_p}$, we must examine the odds ratio, $\frac{P(D | \overline{H_p})}{P(D | H_p)}$. $H_p$ is favoured if this ratio is less than 1, and equation~\ref{eqn:posteqprior} gives us a measure of confidence in this in terms of an actual probability. This is the standard Bayesian approach to testing two hypotheses with equal priors (e.g.\ Gregory~\citealp{gregory_2005}). In contrast, working with only $P(D | \overline{H_p})$ or only $P(D|H_r)$ are examples of what I called incomplete inference.

What does this calculation reveal for for the analyses of Raup \& Sepkoski and Heisler \& Tremaine? It's not
easy to convert their quoted confidence levels into consistent probabilities because they use different methods for detecting periods and for assigning significances. Furthermore, Raup \& Sepkoski~\citep{raup_sepkoski_1986} look for evidence of periodicity at a range of periods, for which there is a higher probability of detecting something significant than at a specific period. So what follows is only approximate.

Raup \& Sepkoski~\citep{raup_sepkoski_1986} identify the 26\,Myr period because it crosses a 99.9\% confidence level in their nonparametric test, implying $p=0.001$.
The confidence for a period at {\em any} period must be lower (they are implicitly using prior information in identifying 26\,Myr with the stated confidence).
They also describe Monte Carlo simulations of 500 random time series (from $H_r$), of which 23 gave rise to a significant ($p<0.001$) period at some value. So really $P(D | \overline{H_p})$ lies somewhere between $0.001$ and $23/500=0.046$ depending on what we are testing.

On the other hand, the simulations by Heisler \& Tremaine \citep{heisler_tremaine_1989} deal with the detectability of a periodic signal at a single frequency.
They find that (only) 55\% of their error-perturbed time series result in this period being detected, implying that $P(D | H_p) = 0.55$.  

Putting these two values, $P(D | \overline{H_p}) = 0.001$ and $P(D | H_p) = 0.55$, into equation~\ref{eqn:posteqprior} gives $P(H_p | D) = 0.998$. 
On this basis, the strong evidence for a period at 26\,Myr outweighs the somewhat modest probability of detecting the period at all. Had we used $P(D | \overline{H_p}) = 0.046$ instead of $0.001$ (which is more realistic, as we are interested in any period), we would get $P(H_p | D) = 0.92$, which is still in favour of a period but much less confident.  More significantly, there are neglected hypotheses ($H_p$ is not the only alternative to $H_r$) so there should be additional terms on the right-hand-side of equation~\ref{eqn:probsum} and thus in the denominator of equation~\ref{eqn:posteqprior}. As these terms are always positive they would decrease $P(H_p | D)$. {\em Neglecting hypotheses leads us to overestimate the confidence in $H_p$.}
As discussed in section~\ref{26myr}, Stigler \& Wagner~\citep{stigler_wagner_1987}
showed that non-periodic models do produce a significant peak with the analysis method of Raup \& Sepkoski, so important alternative models have been neglected and $P(H_p | D)$ is certainly overestimated here.

There is another important issue, namely that of the priors.  We have so far assumed (in equation~\ref{eqn:bayes2}) that the two hypotheses being tested have equal prior probabilities. That is, we have assumed that the unconditional probability of getting a period at 26\,Myr (plus/minus some bin width) is 0.5.  But why would we want to assume a large probability for a specific period {\em before we've even seen the data}? It seems more reasonable to give equal priors to the hypotheses ``periodic'' (for any period) and ``non-periodic''. This must lower the prior for a specific period by a large amount, thus reducing the posterior probability for that period.  Ultimately, the Raup \& Sepkoski analysis is insufficient to provide significant evidence for $H_p$.

There are other subtle issues, and at some point the quality and quantity of the data may not justify a much more thorough analysis.  The main point of this discussion was to convince the reader that assessing evidence for periodicity is not a trivial matter, and is far from being concluded by a low $p$-value for some specific null hypothesis.

%%%%%%%%%%%%%%%%%%%%%%%%%%%%%%%%%%%%%%%%%%%%%%%%%%%%
\section{Terrestrial mechanisms of biological change}\label{terra_mech}

Mass volcanism has occurred many times in Earth's history. This could have had a significant impact on evolution via climate modification (Wignall~\citealp{wignall_2001}). For the first few months after an eruption, SO$_2$ causes local warming via the greenhouse effect, but then reacts with water to produce sulphate aerosols. These, as well as the ash from the eruption, reflect incident sunlight resulting in substantial global cooling. Although the the ash and sulphates rain-out within a few years, they could have longer-term effects via feedback effects (e.g.\ increased snow accumulation at high latitudes during the cool period).  The Mount Pinatubo (Philippines) eruption in 1991 (a VEI 6 event, which occur every 100 years or so) produced sufficient ash to reduce global temperatures by about 0.5\deg\,C. (The Krakatoa eruption in 1883 was of similar magnitude.) The year after the 1815 Mount Tambora (Indonesia) eruption (VEI 7; every 1000 years) was called ``the year without summer'', resulting in crop failure and famine in China, Europe and North America (although a causal connection is disputed).  Other gases released by volcanic eruptions would deplete the ozone layer (Cl$_2$) and precipitate out as acid rain (H$_2$SO$_4$, HCl and HF), both on a 10 year timescale. Volcanos release large amounts of \co2\ which can reside in the atmosphere for up to 100\,000 years. Long after the aerosols have cleared there could have been a long period with a warmer climate.

The largest recorded extinction in the Phanerozoic is at the Permian-Triassic boundary, some 250\,Myr BP.  While the rapidity of extinction among both land and marine organisms suggests an impact cause, the coincidence of mass volcanism as recorded in the Siberian traps (volcanic flood basalts) suggests a terrestrial origin (Erwin~\citealp{erwin_2003}). Indeed, four of the ``big five'' mass extinctions coincide 
very closely with times of mass volcanism (Wignall~\citealp{wignall_2001}, Alvarez~\citealp{alvarez_2003}). Alvarez~\citep{alvarez_2003} argues that beyond this strong temporal correlation there is no direct evidence for a causal connection between volcanism and these mass extinctions, although others argue that some of the key signatures interpreted as evidence of an impact (e.g.\ the iridium layer) can be produced by volcanism (see Glen~\citealp{glen_1994}).

Pandey \& Negi~\citep{pandey_negi_1987} suggest that volcanic activity (as measured by the number of events per unit time) shows a periodic variation over the past 250\,Myr with a period of around 33\,Myr, although this is based on a eyeball analysis of the data. They note that this is close to the Galactic disk plane crossing period. Abbott \& Isley~\citep{abbott_isley_2002} go further, identifying a strong correlation between the terrestrial and lunar impact history and terrestrial mantle plume activity (to a degree which depends on the amount of smoothing applied to the data).  They suggest that impacts may increase the amount of volcanic activity, although their method and conclusions have been criticized by Glikson~\citep{glikson_2003}.  This is not a new suggestion, and while it would be a convenient solution to the volcanism vs.\ impact debate, it seems to have little support.

Various other terrestrial mechanisms have been suggested as causes for mass extinction.  Changes in sea level are often implicated in mass extinctions of marine animals (e.g.\ Hallam~\citealp{hallam_1989}). A drop in the sea level -- either globally due to water becoming locked in ice sheets or due to local uplift -- would reduce the submerged continental shelf area, a region of high biological productivity.  Lower sea levels can also influence the climate through the modified atmosphere--ocean connection.  
Sea level {\em rises} are also implicated in mass extinctions, perhaps more so than drops in the sea level.
On the other hand, Bambach~\citep{bambach_2006} concludes that while low sea levels correlate with extinction, they probably did not cause them.  
There is evidence suggesting there have been long periods of severe oxygen deficiency in the oceans, presumably a result of 
the termination of deep ocean circulation by some mechanism. This has been implicated as the cause of some mass extinction episodes in marine species (Hallam~\citealp{hallam_2004}).
Continental drift and mountain building also affect climate, because the location of land mass influences wind and ocean currents (e.g.\ the uplift of the Tibetan plateau and Himalaya mountains following the collision of India with Asia in the late Cenozoic). A similar effect could come about from very large impacts (diameter $>100$\,km), which some have predicted could significantly alter the Earth's surface (Teterev~\citealp{teterev_etal_2004}).  Finally, the release of methane (a strong greenhouse gas) from submarine methane hydrates has also been suggested. As alternative hypotheses, terrestrial mechanisms must be considered. Indeed, some would suggest that with the possible exception of the K-T boundary, all mass extinctions can be explained by terrestrial mechanisms.
But I will now turn to the main focus of this article: extraterrestrial mechanisms.

%%%%%%%%%%%%%%%%%%%%%%%%%%%%%%%%%%%%%%%%%%%%%%%%%%%%
\section{Extraterrestrial mechanisms of biological change}\label{ET_mechanisms}

There are several extraterrestrial mechanisms which could affect the Earth's biosphere on long timescales.  Some could be triggered as a consequence of the Sun's path through the Galaxy, e.g.\ passages near spiral arms or through the Galactic plane. This is discussed in section~\ref{solar_motion}.  For several of these mechanisms, the postulated immediate cause of extinction is climate change (see also Feulner~\citealp{feulner_2009}).

\subsection{Minor body impacts, Oort cloud perturbation and comet capture}\label{impacts}

There are hundreds of large impact craters on the Earth (Shoemaker~\citealp{shoemaker_1983}).\footnote{See \url{http://www.unb.ca/passc/ImpactDatabase/} for a compilation} It is widely accepted that the mass extinction 65\,Myr ago at the K-T boundary was caused at least in part by an asteroid impact, as evidenced by a global iridium deposit (Alvarez et al.~\citealp{alvarez_etal_1980}) and the identification of the Chicxulub crater in Yucatan, Mexico (Hildebrand et al.~\citealp{hildebrand_etal_1991}).  Some contest this view, and it is possible that some of the extinction was caused by major volcanism which occurred at the time in central India (the Deccan traps). Evidence put forward in support of this is an apparent increase in extinction prior to the impact.  Indeed, there has been a significant debate between proponents of volcanic and impact causes of mass extinctions (Glen~\citealp{glen_1994}).\footnote{Glen~\citep{glen_1994} (esp.\ pp.~68--72) discusses how a rift between groups arose on this point. He describes how different scientific communities have employed very different standards of evidence and how, in some cases, they reached quite different conclusions based on the same data.} Both Hallam~\citep{hallam_2004} and Alvarez~\citep{alvarez_2003} argue that while there is strong evidence for a giant impact having caused the K-T extinction (shocked quartz, tektites, iridium, an appropriate crater), such evidence is lacking for other mass extinctions. 

These considerations aside, it is clear that impacts of large asteroids or comets have occurred many times and can cause widespread extinction. The mechanism is either the violence of the impact itself (blast, fires, earthquakes, tsunamis) or changes in the climate.  Concerning the latter, stratospheric dust and sulfates released by the impact (as well as soot from fires) would remain in the atmosphere for a year or so and result in severe global cooling, a similar consequence to massive volcanic eruptions (see section~\ref{terra_mech}). Carbon may also be injected into the atmosphere, and combined with \co2\ from fires could lead to a longer-term ($10^5$\,yr) global warming. A very large impact could even eject the atmosphere. It has been estimated that the Chicxulub crater (diameter 180\,km) was caused by an impactor with a kinetic energy of $10^8$\,Mt TNT equivalent (1\,Mt TNT = 4.2$\times 10^{15}$\,J) (Toon et al.~\citealp{toon_etal_1997}).  This may have released enough dust to make the atmosphere so opaque that photosynthesis stopped and animals had insufficient light to forage for food.  Assuming this was an asteroid with a typical density of 2500\,\kgm\ and impacted with a relative velocity of 15\,\kms, it would have had a diameter of 10--15\,km. For comparison: the Tunguska object and the impactor at Meteor Crater in Arizona both had a kinetic energy of 10--15\,Mt TNT and diameters of around 50\,m; Krakatoa exploded with about 50\,Mt TNT (Shoemaker~\citealp{shoemaker_1983}, Toon et al.~\citealp{toon_etal_1997}); the Hiroshima nuclear bomb had a yield of 13\,kt TNT (it had an efficiency of just 1\%); the most powerful nuclear weapons tested by the Americans and Soviets had yields of up to 50\,Mt TNT (Garwin \& Charpak~\citealp{garwin_charpak_2001}; the largest weapons in current arsenals are ``only''  around 1\,Mt TNT).
Impacts with an energy of $10^4$--$10^5$\,Mt TNT (corresponding to comets or asteroids about 1\,km in size) become significant on a global scale and are estimated to occur once every 300\,000 years or so (Toon et al.~\citealp{toon_etal_1997}, Chapman~\citealp{chapman_2004}).  

The main sources of potential impactors are Near-Earth Asteroids (such as the Atens, Apollos and Amors) and comets (Shoemaker~\citealp{shoemaker_1983}).  Ongoing surveys for Near-Earth Asteroids have either already detected or will soon detect almost all down to sizes of about 1\,km (e.g.\ Morrison~\citealp{morrison_2003}, Harris~\citealp{harris_2008}). In contrast, the survey completeness down to Tunguska-sized objects (50m, which are estimated to impact once every thousand years or so) is just a few percent.

The major (and perhaps sole) source of long-period comets entering the inner solar system is the Oort cloud. This is composed of minor bodies which orbit the Sun with aphelion distances of around 30\,000-100\,000\,AU (cf.\ 30\,AU for the approximate radius of Neptune's orbit) and are believed to be a remnant of the formation of the solar system. As the size of the Oort cloud is similar in magnitude to the distance to the nearest star (Proxima Centauri, at 270\,000\,AU), the Oort cloud could be perturbed by the passage of nearby stars or Giant Molecular Clouds (GMCs). This could increase the frequency with which comets are kicked into the inner solar system and potentially hit the Earth. 
Such a perturbation would probably release many coments, creating a comet shower in the inner solar system and many impacts on the Earth spread over a few Myr.  This is consistent with claims for some mass extinctions being drawn out over a similar timescale. It could also explain the subsequent discovery that the iridium feature at some extinction boundaries is not a sharp spike but has ``shoulders'', the finite width of the feature reflecting the stochastic distribution of impacts (Glen~\citealp{glen_1994}, p.~70).  Even without an actual impact, it has also been suggested that dust from comets could enter the Earth's atmosphere and affect its climate (Hoyle \& Wickramasinghe~\citealp{hoyle_1978}, Torbett~\citealp{torbett_1989}, Shaviv~\citealp{shaviv_2003}).

Given that the stellar density increases towards the Galactic midplane and to a lesser extent inside spiral arms, both have been suggested as triggers for the perturbations (Napier \& Clube ~\citealp{napier_clube_1979}, Napier~\citealp{napier_1998}).  The perturbation could also kick comets away from the Sun, and as the other stellar system presumably has its own Oort cloud, our Sun may capture its comets (Clube \& Napier~\citealp{clube_napier_1982a}). (This has implications for the interpretation of dates of impact material in the solar system: it may have come from another stellar system.) Clube \& Napier~\citep{clube_napier_1982b} speculate that large impacts change the angular momentum of the Earth's core and mantle which could in turn trigger geomagnetic reversals and plate tectonic activity. In other words, different phenomena as recorded in the geological record may have a causal connection, which may in turn be related to mass extinctions or climate change.

Heisler \& Tremaine~\citep{heisler_tremaine_1989} have suggested that the Galactic tide is probably a more significant source of gravitational perturbations than the passage of GMCs, in which case we would expect no relation between impacts and disk plane crossing or spiral arm passages.  Wickramasinghe \& Napier~\citep{wick_napier_2008} estimated the flux of comets due to perturbations from the Galactic tide and molecular clouds. They find that the flux increases about an order of magnitude above the background rate on timescales of 25--35\,Myr, which is consistent with what they describe as a weak periodicity in the cratering record of 36\,Myr.

The perturbation/impact mechanism is certainly a plausible one for causing mass extinctions.  The relevance depends on the number and mass of comets in the Oort cloud as well as the size and frequency of the perturbing effects.  Large impacts could cause significant devastation and wipe out species, but whether they actually {\em would} depends on the complex reaction of ecosystems. The climate would recover relatively quickly (decades, unless significant amounts of \co2\ are released) so the effect on the biosphere is an impulse on geological timescales.

\subsection{Cosmic rays}\label{cosmic_rays}

It has been suggested that Galactic cosmic rays could have an impact on the Earth's climate via cloud formation (e.g.\ Shaviv \& Veizer~\citealp{shaviv_veizer_2003}, Carslaw et al.~\citealp{carslaw_etal_2002}, Shaviv~\citealp{shaviv_2005}, Kirkby~\citealp{kirkby_2007}).  The basic argument is: (1) cosmic rays cause ionization in the troposphere; (2) these ions act as nucleation sites for water droplets which form clouds; (3) low altitude clouds contribute a net negative radiative forcing (cooling). Hence an increased cosmic ray flux would cause global cooling. 

The cosmic ray--cloud/climate mechanism has many uncertainties.  While cosmic rays are an important source of ionization in the atmosphere, it is not yet clear whether they are an important source of nucleation compared to neutral molecules (e.g.\ J{\o}rgensen \& Hansen~\citealp{jorgensen_hansen_2000}). Even if they are, these nuclei must first grow (by a factor of a million in volume) by condensation and coagulation into cloud condensation nuclei (CCN) before becoming a source of cloud formation. The mechanism by which this growth proceeds remains uncertain.
Partly for this reason, it has not yet been established whether the observed amplitudes of variation in the cosmic ray flux induces sufficient variation in the CCN density (Kirkby~\citealp{kirkby_2007}). 
Even if cloud drop nucleation does increase with cosmic ray flux, this does not necessarily translate into a larger areal coverage of clouds: it could rather increase their height or optical depth (J{\o}rgensen \& Hansen~\citealp{jorgensen_hansen_2000}, Carslaw et al.~\citealp{carslaw_etal_2002}).  Another important point is that clouds have both a cooling effect (by reflecting sunlight) and a warming one (by reradiating thermal radiation back to the Earth). Although it seems likely that low altitude clouds ($<3$\,km) produce a net negative radiative forcing (e.g.\ IPCC~\citealp{ipcc_2007} section 8.6.3.2, Kirkby~\citealp{kirkby_2007}), they will only contribute a net cooling when they occur over land or sea which has a lower albedo than the cloud. Over arctic regions, high albedo snow and ice provide a strong cooling by reflecting sunlight. Low altitude cloud cover here will reduce this effect and thus contribute a net warming.

Possible cloud formation is not the only impact of cosmic rays. The ions generated could set up a global atmospheric electric current which itself may have other atmospheric effects (Carslaw et al.~\citealp{carslaw_etal_2002}).  
Furthermore, the muons created by high energy cosmic rays from supernovae or gamma-ray bursts could kill organisms directly or damage their DNA.

The Earth is exposed to a continuous flux of Galactic cosmic rays (with energies of a few to a few tens of GeV), a large portion of which is believed to originate in shock fronts in supernova remnants (e.g.\ Lockwood~\citealp{lockwood_2005}).
The cosmic ray flux reaching the Earth is modulated by solar activity via the interaction of the solar wind with the Earth's magnetosphere (section~\ref{sun}). Thus solar variability on a timescale of years is a potential mechanism for cosmic ray induced climate variability. Cosmic rays are also emitted from the Sun itself, generated at the shock fronts of explosive events on the Sun's surface (e.g.\ flares, coronal mass ejections) and typically have energies below 1\,GeV.
In addition, a nearby supernova would generate a large and potentially lethal burst of cosmic rays (section~\ref{sne}).  
The progenitors of core-collapse supernovae are short-lived, massive stars in star forming regions. As these (and their remnants) are concentrated towards the Galactic plane and in spiral arms, this has motivated some researchers to look for evidence of a correlation between climate/extinction and the solar motion on timescales of tens to hundreds of million years (see section~\ref{solar_motion}).

What evidence is there for cosmic rays affecting climate on geological timescales?  Shaviv \& Veizer~\citep{shaviv_veizer_2003} compare a \oproxy\ temperature proxy constructed by Veizer et al.~\citep{veizer_etal_1999} over the Phanerozic with the cosmic ray flux inferred by Shaviv~\citep{shaviv_2003} from meteorites and three atmospheric \co2\ proxies. They find that the cosmic ray flux -- but not the \co2\ level -- correlates with the temperature record and so conclude that \co2\ has much less effect on global temperatures than other research has shown.  Rahmstorf et al.~\citep{rahmstorf_etal_2004} refute this result on three grounds. First, they claim that the purported correlation is largely a result of several arbitrary ``adjustments'' to the data. Second, they note the tenuousness of the adopted method of inferring variations in the cosmic ray flux from meteorites (see section~\ref{spiral_arms}). Third, they report work by Royer et al.~\citep{royer_etal_2004} which shows that the temperature calibration of the \oproxy\ proxy from Veizer et al.~\citep{veizer_etal_1999} must be corrected for sea water pH. When this is done, the correlation between \oproxy\ and cosmic ray flux vanishes.  Royer et al.~\citep{royer_etal_2004} acknowledge that cosmic rays may have some influence on climate, but that this is probably minor on multimillion year timescales compared to the effect of \co2.

The cosmic ray--climate link is particularly controversial because some researchers have claimed it explains a significant part of post-industrial global warming. The palaeontological record and a supposed influence on it by astronomical phenomena have been adopted to support this by claiming that {\em if} cosmic rays are relevant to climate on hundred million year timescales {\em then} they must be relevant on decadal timescales. Although this is a logical non sequitur -- we must at least consider the amplitudes of the effects -- it is still relevant to ask what evidence there is for a link on other timescales.

A specific claim is that cloud cover over the past few decades correlates well with both the cosmic ray flux and the solar activity (e.g.\ Friis-Christensen \& Lassen~\citealp{friis_lassen_1991}, Svensmark \& Friis-Christensen~\citealp{svensmark_friis_1997}, Marsh \& Svensmark~\citealp{marsh_svensmark_2000}).  The idea is that solar activity affects the solar wind, which in turn provides the Earth with some shielding against Galactic cosmic rays. Those reaching the Earth interact further with the Earth's magnetosphere, resulting in a latitude-dependence of the cosmic ray flux. Many of these claims have been rebutted (e.g.\ J{\o}rgensen \& Hansen~\citealp{jorgensen_hansen_2000}, Laut~\citealp{laut_2003}, Damon \& Laut~\citealp{damon_laut_2004}, Sloan \& Wolfendale~\citealp{sloan_wolfendale_2008}).  For example, the data and data analysis supporting the claims of Friis-Christensen \& Lassen~\citep{friis_lassen_1991} and Svensmark \& Friis-Christensen~\citep{svensmark_friis_1997} have been strongly criticized by Laut~\citep{laut_2003} and Damon \& Laut~\citep{damon_laut_2004}. After correcting for apparent flaws in the methodology, they show that there is no link between cloud cover and cosmic rays. 
J{\o}rgensen \& Hansen~\citep{jorgensen_hansen_2000} note that the reported correlation is relatively weak and that better established mechanisms of observed events provide a more plausible explanation (e.g.\ El Ni\~no--Southern Oscillation, volcanism).  Damon \& Laut~\citep{damon_laut_2004} further report on new data covering 1992--2003 which confirm no correlation between cosmic ray flux and {\em total} (all altitudes) global cloud cover. Furthermore, before 1994 changes in the cloud cover lag behind changes in the cosmic ray flux by six months, whereas from what we know of the cosmic ray mechanism the lag should be not much more than one day (Laut~\citealp{laut_2003}).

Using two independent estimates of low altitude cloud cover from the International Satellite Cloud Climatology Project over the period 1983--1999, Kristj\'ansson et al.~\citep{kristjansson_etal_2002} show that there is either no significant correlation or even a negative correlation between cosmic rays and low altitude cloud cover.  On the other hand, Pall\'e \& Butler~\citep{palle_butler_2002} summarize arguments for and against a cosmic ray--cloud link over the past 50--120 years and conclude that a lack of quality data does not allow us to ``totally dismiss the link between [Galactic cosmic rays] and cloudiness''.

Erlykin et al.~\citep{erlykin_eta_2009} show that there is a common variation between the cosmic ray flux in the Earth's atmosphere, sunspot number, solar irradiance and global average surface temperature over the past 50 years (1956--2002) which varies on a timescale of twice the 11-year solar cycle. (Only two ``cycles'' are seen so we cannot call this ``periodic''. The analysis uses an 11-year moving average centered on the point of interest. While this smooths shorter timescale varations, it does not eradicate them and, importantly, by using a symmetric smoothing window no phase shift is introduced.) The temperature, solar irradiance and sunspot number variations are in phase, whereas the cosmic ray-flux lags behind by 2--4 years, so it cannot be a cause of the temperature variation. 
They further estimate that the direct impact of cosmic rays on the radiative forcing of the Earth is less than $+0.07$\deg\,C since 1956, so contributing less than 14\% of global warming.

In summary, there is no strong evidence that cosmic rays have significantly influenced climate on geological timescales.  There is now a broad consensus among scientists that, although cosmic rays may have an effect on cloud microphysics (via a mechanism only poorly understood) and thus on climate, the data show that they had at most a minor influence on post-industrial global warming (IPCC~\citealp{ipcc_2007}).  This is further reinforced by the fact that we have much more support for an alternative hypothesis for climate change, namely \co2\ and other greenhouse gases. This consensus is not always reflected proportionally in the media.

\subsection{Supernovae and Gamma-Ray Bursts}\label{sne}

Supernovae release large amounts of energy in the form of hard x-rays and cosmic rays.  These could cause widespread extinction through at least three mechanisms. First, the radiation can kill organisms (on the hemisphere facing the blast) via direct cell destruction or damage to DNA. Second, ionizing radiation creates nitric oxide (NO) in the upper atmosphere, destroying ozone. A single blast could leave the ozone layer depleted for hundreds of years, exposing life to harmful solar UV radiation which can damage DNA even in water at depths of several meters (Ruderman~\citealp{ruderman_1974}).  Third, \no2\ formed from the NO is a strong absorber of visible radiation from the Sun leading to global cooling (Thomas et al.~\citealp{thomas_etal_2005}). 
As just discussed (section~\ref{cosmic_rays}), cosmic rays may also affect climate via cloud formation. Tanaka~\citep{tanaka_2006} estimates that a supernova within 12--15\,pc would increase the flux of 10--100\,GeV cosmic rays by a factor of 4--8, although 
how this translates to condensation nuclei is unclear.  Apart from the initial blast, cosmic rays are also emitted from the supernova remnant for millions of years thereafter by shock wave processes.

Ellis \& Schramm~\citep{ellis_schramm_1995} examined the possibility that nearby supernovae could have caused mass extinctions on Earth.  Using a supernova rate of 0.1 per year in our Galaxy and an average stellar density of 1\,pc$^{-3}$, they estimate that a supernova would occur within 10\,pc of the Sun every 240\,Myr or so (this is only an order of magnitude estimate). 
Assuming that core-collapse of massive stars are the dominant cause of Galactic supernovae, then the probability of a nearby supernova increases as the solar system crosses spiral arms, although not by a lot because the space density of supernovae may not be significantly larger in spiral arms (see section~\ref{spiral_arms}).There will also be a change in the local space density of massive stars (and thus supernovae) as the Sun oscillates vertically through the Galactic plane. The magnitude of variation depends on the scale height for massive stars and the amplitude of the motion. It we adopt 100\,pc for the former with an exponential profile (Da-li \& Zi~\citealp{dali_zi_2008}) and 70\,pc for the latter (Gies \& Helsel~\citealp{gies_helsel_2005}), then the maximum increase in density is only $\exp(70/100) = 2$.

Gamma-ray bursts would have a similar effect on the Earth as supernovae, but could be effective out to distances of several kpc. Thomas et al.~\citep{thomas_etal_2005} have modelled their impact on the atmosphere and biosphere is some detail. Melott et al.~\citep{melott_etal_2004} suggest that gamma-ray bursts occur at a rate which could cause two or more mass extinction on the Earth every billion years, and single out in particular the late Ordovician event.

\subsection{Solar variability}\label{sun}

The total solar radiation reaching the top of the Earth's atmosphere (the solar irradiance) is 1367\,W\,m$^{-2}$ averaged over the orbit.  Convection and magnetic activity in the Sun's atmosphere (Sun spots) result in a variation of the solar irradiance by 0.1\% over the 11 year solar cycle.  This amplitude is too small to cause the ice ages on the Myr timescale discussed earlier.  Lockwood~\citep{lockwood_2005} states that periodic variations in the Earth's climate on timescales of a decade or less are mostly smoothed out by the atmosphere--ocean coupling and large thermal capacity of the oceans.  (This would therefore include the annual 7\% variation in solar irradiance due to the eccentricity of the Earth's orbit.) 
However, UV and shorter wavelength variations are larger (Shaviv~\citealp{shaviv_2003}). Although UV is mostly absorbed in the stratosphere, the variations could be propagated down through the atmosphere.  Furthermore, the intensity of spots varies on longer cycles, and longer durations with fewer spots (such as the Maunder Minimum around 1645--1715) have been associated with climate change.  According to Lockwood \& Fr\"ohlich~\citep{lockwood_froehlich_2007}, the evidence suggests that solar variability has had an impact on climate over the past few centuries.

In addition to variations in the electromagnetic flux, there are variations in the cosmic rays (solar protons and electrons) emitted from the Sun. These charged particles form the solar wind and are responsible for the heliosphere, which modulates the flux of Galactic cosmic rays reaching the Earth. An increase in the solar activity ``strengthens'' the heliosphere and so lowers the cosmic ray flux reaching the Earth. The cloud mechanisms described in section~\ref{cosmic_rays} have been invoked by some authors to provide the connection between solar activity and Earth climate.  Lockwood \& Fr\"ohlich~\citep{lockwood_froehlich_2007} show that data on the potentially relevant phenomena of the Sun -- total solar irradiance, solar magnetic flux, cosmic ray flux from neutron counts, sun spot number -- since 1985 show variations in the direction {\em opposite} to that required for them to be responsible for the recent rise in global temperatures.
According to G.\ Feulner (private communication, April 2009), 
terrestrial temperature variations over the 11 year solar cycle are detectable, and, moreover, that they
can be reproduced within climate models by the variation in the solar irradiance alone, without having to invoke effects of cosmic rays on the cloud cover.

\subsection{Variations in the Earth's orbit about the Sun}\label{orbit}

The orbit of the Earth about the Sun is nominally an ellipse, but it is perturbed by the gravitational force of other bodies in the solar system. This causes both the inclination of the Earth's orbital plane and the eccentricity of its orbit to vary.  The Sun and Moon also impart torques on the Earth which cause the Earth's spin axis to precess and its obliquity (angle with respect to the orbital axis) to vary.

These perturbations have been modelled very accurately using classical mechanics.  By expressing them as a series expansion we can identify terms with different periods and amplitudes.  The eccentricity of the Earth's orbit varies between almost 0 and 0.05 (it is currently 0.017 and decreasing) with dominant periods of 95, 125 and 400\, kyr with relative amplitudes of 1.2, 1.0 and 1.7 respectively. Together these account for 90\% of the signal.  (These and other figures in this paragraph are taken from Muller \& MacDonald~\citealp{muller_macdonald_2000}.) The average distance of the Earth from the Sun depends on the eccentricity, so variations in this translate into variations in the average annual solar irradiance.\footnote{If $a$ is the semi-major axis and $e$ the eccentricity, the time-averaged distance (averaging over the mean anomaly or phase of the orbit) is $a(1 + e^2/2)$.}  In contrast, variations in precession, obliquity and orbital inclination only affect the geographical distribution of the flux. These can nonetheless induce an ice age by preventing winter ice from melting in the summer, and
subsequent changes in tree cover and sea ice enhance this via a positive feeback
 (IPCC~\citealp{ipcc_2007}, chapter 6).  
The precession period of the Earth's spin axis is 25.8\,kyr, but as this is modulated by eccentricity variations, the precession parameter (which is relevant to variation of the solar irradiance) shows periods at 19, 22 and 24\,kyr. The axis of the Earth's orbital plane also precesses. The orbital inclination varies between 0\deg\ and 4\deg\ at a dominant period of about 70\,kyr.  Because the directions of both the orbital axis and the spin axis vary with different periods and amplitudes, then the angle between them (the obliquity) varies too. The result is that the extent of the tropics varies between 22.1 and 24.5\deg\ (it is currently 23.5\deg\ and decreasing) with a dominant period of 41\,kyr (and weaker periods at 29 and 53\,kyr).

The similarity between periods in the climate record over the past 3\,Myr (discussed in section~\ref{climate_myr}) and periods in the perturbation terms of the Earth's orbit over a similar timescale, has lead many to suggest a causal connection (e.g.\ Hays et al.~\citealp{hays_etal_1976}). This is sometimes called the Milankovitch theory of the ice ages, although others proposed it before him and others have modified it since. Its broad formulation is generally accepted to explain the occurrence of recent ice ages (e.g.\ IPCC~\citealp{ipcc_2007}), although some of the details are debated. Part of the evidence comes in the form of very narrow peaks in the climate power spectrum. Narrow peaks imply a non-dissipative process (Muller \& MacDonald~\citealp{muller_macdonald_2000}). This in turn is taken to imply a mechanism which is similarly non-dissipative, such as planetary orbits. One may argue that the simultaneity of ice ages in the northern and southern hemispheres argues against this, but because the northern hemisphere possesses two thirds of the terrestrial land surface, it is the northern hemisphere insolation which is responsible for triggering the ice ages.

In summary, while Earth axis and orbit variations may well explain some climate change on timescales of a few tens to a few hundreds of kyr, they seem not to offer an explanation for change or periodicity in climate on longer timescales.

\subsection{Other mechanisms}

Encounters of the solar system with interstellar clouds could produce global cooling as dust from the cloud lowers the solar irradiation. Shaviv~\citep{shaviv_2003} has coupled this with the idea that a bow shock from the cloud reduces the size of the heliosphere below 1\,AU and thus exposes the Earth to a greater flux of Galactic cosmic rays. 
Contrarily, Hoyle \& Lyttleton~\citep{hoyle_lyttleton_1939}
suggested that matter from an interstellar cloud falling into the Sun could raise the Sun's luminosity (via the release of gravitational energy) and that this in turn could trigger an ice age via increased precipitation.
McCrea~\citep{mccrea_1975} took up this idea and suggested that sufficient matter cloud be provided by the Sun's passage through dust lanes at the edge of spiral arms.

Another suggestion is that the Sun is in a wide binary system, with a faint M dwarf or later-type companion in a long period orbit (Davis et al.~\citealp{davis_etal_1984}; Whitmire \& Jackson~\citealp{whitmire_jackson_1984}).  If the companion (colourfully named ``Nemesis'' in the article of Davis et al.~\citealp{davis_etal_1984}) is in a highly elliptical orbit, then perihelion passages at around 30\,000\,AU would be close enough to perturb the Oort cloud and eject large numbers of comets toward the inner solar system (section~\ref{impacts}).  This idea was originally proposed as a mechanism to explain the claimed 26\,Myr periodicity in extinction from Raup \& Sepkoski~\citep{raup_sepkoski_1984}. 
To achieve an orbital period of this order, then with the required eccentricity of around 0.7 (or more) the companion would need a semi-major axis of order $10^5$\,AU.
Davis et al.~\citep{davis_etal_1984} estimated that such a companion could perturb the orbits of some $10^9$ Oort comets, of which 25 would hit the Earth. However, the semi-major axis for the proposed companion is unusually large for binary systems (0.5\,pc), leading some authors to suggest that the Galactic tide or close encounters with stars or interstellar clouds would unbind the system on a timescale of a Gyr (Torbett \& Smoluchowski~\citealp{torbett_smo_1984}).  Even if it remains bound, these perturbations are likely to make the orbit unstable: One calculation predicts that a 26\,Myr orbital period would vary by 10--20\% over 250\,Myr (Hut~\citealp{hut_1984}). Interestingly, this could actually speak in favour of this mechanism if extinctions are demonstrated to be quasi-periodic (rather than strictly periodic or non-periodic). No candidate for Nemesis has been found in deep, all-sky surveys.

%%%%%%%%%%%%%%%%%%%%%%%%%%%%%%%%%%%%%%%%%%%%%%%%%%%%
\section{Solar motion through the Galaxy}\label{solar_motion}

The orbit of the Sun through the Galaxy can be reconstructed from knowledge of (1) the gravitational potential of the Galaxy and (2) the present position and velocity of the Sun, using numerical integration.  To determine the gravitational potential one must specify a mass model for the Galaxy which can then be fit using stellar kinematic data (e.g. Dehnen \& Binney~\citealp{dehnen_binney_1998a}).  These stellar velocities in turn are derived from astrometry (positions, parallaxes and proper motions -- five components of the six-dimensional phase space vector) and radial velocities (the sixth component).  The solar motion is likewise determined from the kinematics of stars in the local neighbourhood and the adoption of a model for the Galactic rotation (e.g.\ Dehnen \& Binney \citealp{dehnen_binney_1998b}, Fuchs et al.~\citealp{fuchs_etal_2009}).  
The position of the Sun relative to the Galactic plane may be inferred from the distribution of populations of disk stars. The distance to the Galactic centre may be determined by radio mapping of the gas in the Galactic disk or from the distances to objects believed to be distributed symmetrically about the Galactic centre (e.g.\ RR Lyrae stars and globular clusters).

The Sun currently moves on an approximately circular orbit in the disk plane of the Galaxy.  Estimates of the distance of the Sun from the Galactic centre ($R_0$) published since 1974 range from 6.7--9.6\,kpc.  Most estimates lie in the range 7.5--8.5\,kpc with 8.2\,kpc often being taken as a best estimate (Perryman~\citealp{perryman_2009}, section 9.2).
Adopting a rotation speed of the local standard of rest of 220\,\kms, this corresponds to a rotation period around the Galactic centre of 235\,Myr. Yet estimates for this rotation speed vary from 195--255\,\kms, so the period could be anything from 200--265\,Myr (assuming $R_0 = 8.2$\,kpc).  The solar motion is neither perfectly circular nor exactly planar.  The Sun is currently north of the midplane and moving away from it at $7 \pm 1$\,\kms, and is moving toward from the Galactic centre at $9 \pm 1$\,\kms\ (Fuchs et al.~\citealp{fuchs_etal_2009}). Estimates for the current distance of the Sun from the midplane range from 8--35\,pc (Perryman~\citealp{perryman_2009}), depending partly on the population of stars considered. For example, Da-li \& Zi~\citep{dali_zi_2008} estimate $15.2 \pm 7.3$\,pc when using OB stars and $3.5 \pm 5.4$\,pc when using HB stars.

\begin{figure}[htb]
\begin{center}
\includegraphics[scale=0.50, angle=0]{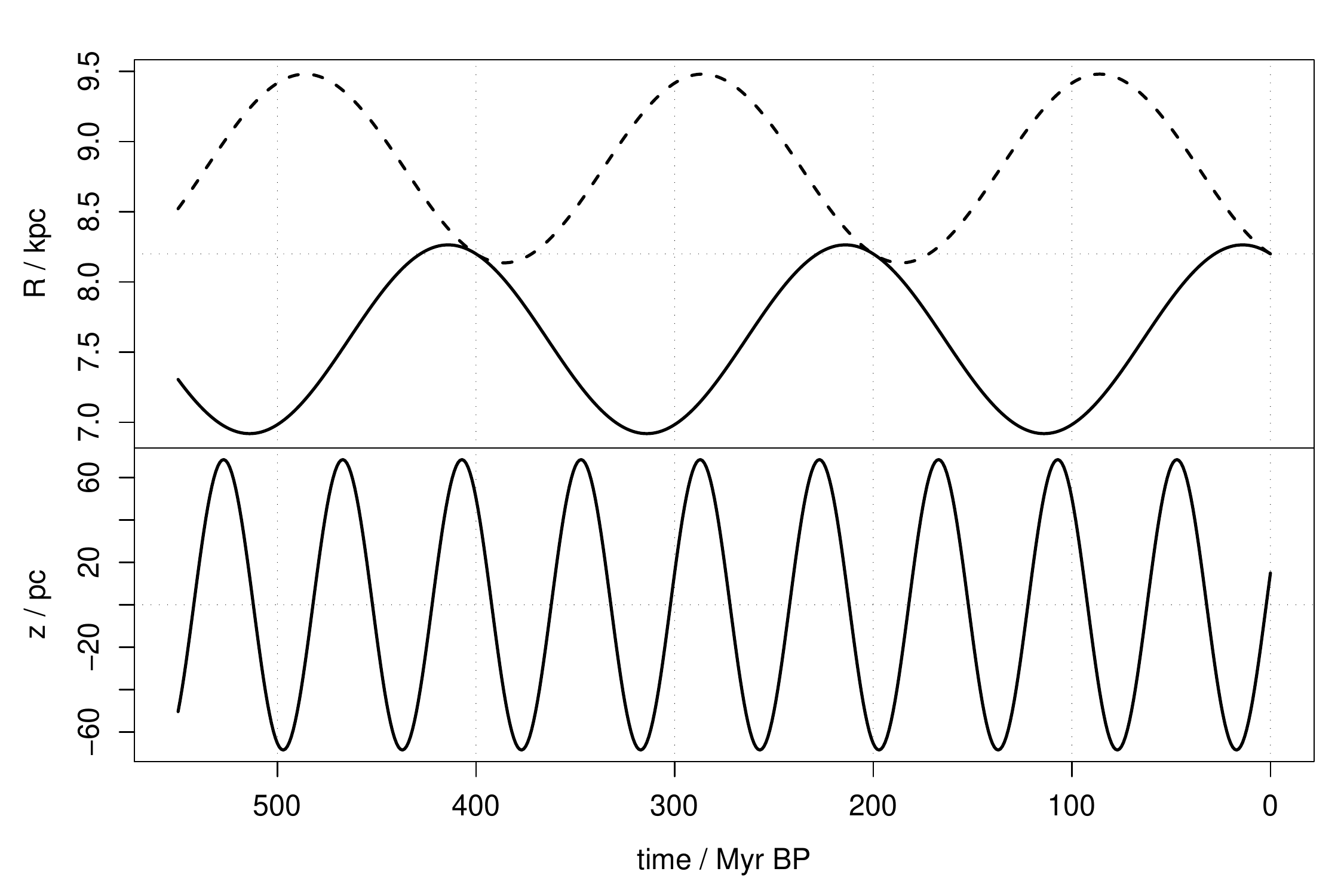}
\caption{Example of the radial (top panel) and vertical (bottom panel) motion of the Sun about the Galactic centre using the sinusoidal epicyclic approximation. The radial motion assumes a period of 200\,Myr, $R_0=8.2$\,kpc, a half peak-to-peak amplitude of 0.67\,kpc and a current solar radial velocity of 9\,\kms\ toward the Galactic centre, giving rise to two solutions (solid and dashed lines). The vertical motion assumes a period of 60\,Myr, 
and the amplitude has been solved for (68\,pc) using values for the 
current position of the Sun (15\,pc north of the plane) and its current vertical velocity (7\,\kms\ to the north). The Galactic midplane is at z=0. It is well established that the Sun lies north of the midplane and the signs of its velocities are well known, but there is significant uncertainty in the values of all the parameters (the values chosen are somewhat arbitrary.)
}
\label{fig:solar_path}
\end{center}
\end{figure}

The perturbations of the Sun's motion about a circular orbit can be described using two independent sinusoidal components in the radial and vertical directions by adopting the epicyclic approximation.  The resulting simple harmonic motion is shown in Figure~\ref{fig:solar_path}. 
As an example of this model, Shuter \& Klatt~\citep{shuter_klatt_1986} took $R_0 = 8.5$\,kpc and used a value of $\theta_0 = 220$\,\kms\ for the circular velocity of the epicentre. Adopting an axisymmetric model of the potential, they derive the period and (half peak-to-peak) amplitude of the two components to be: radial, period\,$=180$\,Myr, amplitude\,$=0.7$\,kpc; vertical, period\,$=66.2 \pm 3.4$\,Myr, amplitude\,=$100$\,pc.
The radial displacement varies from $0.992 R_0$ to $1.156 R_0$.  
Because the disk potential is modelled to drop off exponentially with distance from the Galactic centre, 
this radial motion causes the gravitational potential experienced by the Sun, and therefore its
vertical oscillation period, to change.  The magnitude of this period variation depends upon both the mass gradient in the disk and the size of the radial variations in the solar orbit. These are not well determined, but Shuter \& Klatt estimate the vertical period to have been on average 8\% larger over the past 250\,Myr than the current value.
The implication is that {\em if} extinction events and other phenomena on the Earth are influenced by the vertical motion of the Sun then we should not expect those events to show a constant period.
In their model, Shuter \& Klatt~\citep{shuter_klatt_1986} show that by adopting a constant (best-fit) period of 66.2\,Myr, one will accumulate a phase shift of 21\,Myr over a duration of 250\,My, a third of the mean vertical oscillation period.

Many authors have used this vertical motion or spiral arm crossings in combination with one of the mechanisms outlined in section~\ref{ET_mechanisms} to account for biodiversity variations or climate change. In the next two sections I will examine these claims and the evidence for and against them in more detail.

\subsection{Motion perpendicular to the Galactic plane}\label{z_motion}

There have been many attempts to model the motion of the Sun perpendicular to the plane.  Using a model for the gravitational potential of the disk, Bahcall \& Bahcall~\citep{bahcall_bahcall_1985} derive a quasi-harmonic motion with a period ranging from 52--74\,Myr, the range reflecting uncertainties in both the distribution of dark matter (which dominates) as well as the current vertical position and velocity of the Sun.
The maximum displacement from the plane varies from 49 to 93\,pc, with the motion being significantly non-harmonic beyond about 40\,pc (because not all the matter is concentrated in the plane, but rather falls off exponentially).  

Svensmark~\citep{svensmark_2006} takes a fundamentally different approach. Instead of 
fitting the motion of the Sun using astronomical data, he fits it to \oproxy\ temperature proxy data (section~\ref{geobiodata}) from the past 200\,Myr, using the ad hoc assumption that the temperature of the Earth, $T$, varies as the square of the distance, $z$, of the Sun from the Galactic midplane. Thus \oproxy\ becomes a proxy for $z^2$.
Adopting the harmonic oscillator model for the $z$ motion, $T$ then varies with twice the oscillation frequency.
As his \oproxy\ sample shows a period of around 30\,Myr, this translates to a vertical oscillation period of 60\,Myr. (The disk potential model allows for two perturbations due to two spiral arm crossings, so the vertical period is not strictly periodic. See section~\ref{spiral_arms})

There is a methodological problem with this approach, however, because it {\em assumes} that the Earth gets hotter the further it is from the Galactic plane. This, in turn, is based on the assumption that the mean global temperature is controlled by cosmic rays via cloud cover, as discussed in section~\ref{cosmic_rays}.  But because the solar motion is derived from the \oproxy\ (temperature) data, we cannot then use this to claim that the solar motion fit provides support for the cosmic ray model.  We could only do this if we test the assumption $T \propto z^2$ by comparing the \oproxy\ data with an {\em independent} determination of the solar motion. Otherwise we have circular reasoning. Note also that, because the assumption is not based on a physical model -- it is chosen as the simplest form which gives the required symmetry of $T(z)$ -- $z$ amplitudes cannot be derived from \oproxy.  It should also be mentioned that the phase of the fit (i.e.\ the current $z$ coordinate of the Sun, taken as $9 \pm 4$\,pc north of the plane) is an input to the model, not a prediction.  The most we can conclude is that the selected \oproxy\ data appear to show a period which is of order half that of {\em other} determinations of the solar vertical period.
But this study does not really lend support to the idea that an increased flux of cosmic rays from the Galactic plane causes a net cooling of the Earth. 

Svensmark is by no means the only author to use geological data to try and infer solar motion. There were several efforts in the 1980s, such as the work of Shuter \& Klatt~\citep{shuter_klatt_1986}, to try to constrain the dark matter content of the disk in this way (see also Bahcall \& Bahcall~\citealp{bahcall_bahcall_1985} and Wickramasinghe \& Napier~\citealp{wick_napier_2008}).  Given the significant uncertainties in the geological record, plus the considerable doubt about the existence of (stable) periods, this is a rather suspect approach.

Medvedev \& Melott~\citep{medvedev_melott_2007} also constructed an astronomical model in which extinctions are caused by cosmic rays. They are motivated by the apparent $62 \pm 3$\,Myr period in biodiversity variation from Rohde \& Muller~\citep{rohde_muller_2005}, discussed in section~\ref{phan_periods}.  This is close to the period of the solar vertical oscillations in many dynamical models
and thus {\em twice} the period of Galactic plane crossings. 
If there is a causal connection, then extinctions cannot be associated with star forming regions, spiral arms or anything else concentrated in the plane, but would have to be associated with some plane asymmetry.  They note that the maxima in Rohde \& Muller's analysis coincide with the Sun being near to its northern-most displacement, which is the direction to the Virgo supercluster. Based on these considerations, they propose a model in which a Galactic bow shock is produced by the Galactic wind and the motion of the Galaxy towards Virgo.  This shock is a source of cosmic rays.  The Galactic magnetic field shields the Sun from these to an extent which varies with distance from the Galactic midplane: there is considerably less shielding (perhaps five times the cosmic ray flux on the Earth) when the Sun is at its northern-most displacement.

They test this using a model of the solar motion from Gies \& Helsel~\citep{gies_helsel_2005} to predict the cosmic ray flux at the Earth over the Phanerozoic and compare it with Rohde \& Muller's~\citep{rohde_muller_2005} results.  Minima in diversity phase well with predicted cosmic ray maxima, perhaps implying that cosmic ray-induced cooling causes mass extinctions (although they don't pinpoint a specific extinction mechanism).  Although the data interpretation involves many assumptions and the model is rather sketchy, the authors argue that the 62\,Myr period in the biodiversity data demands an extragalactic explanation.

Many other papers have invoked plane crossings to explain mass extinctions or climate variations with periods around 25--33\,Myr (e.g.\ Raup \& Sepkoski \citealp{raup_sepkoski_1984}, \citealp{raup_sepkoski_1986}, Rampino \& Caldeira~\citealp{rampino_caldeira_1992}, Napier~\citealp{napier_1998}) as discussed in sections~\ref{26myr} and~\ref{geoperiods}.  Two frequently invoked triggers are an increased supernova rate (section~\ref{sne}) and increased rate of comet impacts (section~\ref{impacts}).  However, the increases may not be that large. For example, the local supernova rate at midplane is perhaps only double that at maximum distance from the plane (section~\ref{sne}), so the exposure to cosmic rays (for example) from supernovae and their remnants would also only change by this amount. Whether this is a {\em physically} important difference depends on the details of the climate change or extinction model.
Some authors have proposed a connection between the impact cratering record, mass extinctions and Galactic plane crossings (section~\ref{craters}), but one claimed period of around 13\,Myr for cratering does not fit here. 

While the estimates of the parameters of the solar orbit have a wide range, there is a reasonable consensus that the Sun is currently near the midplane ($z \simeq 15$\,pc compared to a vertical oscillation amplitude of around 70\,pc).
The Sun therefore recently underwent a midplane passage, yet the lack of evidence for a major extinction within the past 15\,Myr has led some to suggest that plane crossings cannot be the cause of mass extinctions (e.g.\ Leitch \& Vasisht~\citealp{leitch_vasisht_1998}).  We could argue that not every midplane crossing will cause a mass extinction (or ice age), but when we permit ourselves to pick and choose we cannot look for periodic phenomena, and indeed could fit almost any model to the geological data. 

In the discussions of sections~\ref{ts_evidence} and~\ref{ts_fun}, I concluded that there is no good evidence for periodicities in climate or mass extinction in the range 25--33\,Myr.  This seems to rule out a relevant periodic influence of Galactic plane crossings, and other studies looking for a specific connection have not provided convincing evidence of a causal link.  The recent 62\,Myr period in biodiversity appears to be on a stronger footing, yet its identification is also subject to the limitation of ``incomplete inference'' discussed in section~\ref{h_testing}, and is not without criticism (see section~\ref{reliability}). It remains an interesting suggestion to be explored further in other data sets.  But given that the periodicity is only significant over the interval 520--150\,Myr BP, a connection to the solar $z$-motion is not obvious. The link to the uncertain cosmic ray--cloud connection and even more uncertain bow shock model are additional, unresolved steps in the causal chain.

\subsection{Spiral arm crossings}\label{spiral_arms}

Spiral arms are sites of increased star formation in the Galactic disk, characterized in external galaxies by the excess of massive young blue stars and giant molecular clouds. The origin of spiral arms is debated. The classical model originating in the work of Lindblad~\citep{lindblad_1938} and Lin \& Shu~\citep{lin_shu_1964} explains them as gravitational in origin, whereby they can be understood as density waves propagating through the disk. In this model the spiral pattern is fixed and rotates as a rigid structure with angular velocity \omp\ (the pattern speed). In contrast, stars and gas in the disk rotate differentially, so they will pass in and out of the arms even if on exact circular orbits. 
Within the corotation radius the stars overtake the spiral arms 
and beyond it are overtaken by the arms.
Only at the corotation radius do stars on circular orbits not move relative to the spiral pattern. The position of this radius is disputed, and various studies have placed it at Galactocentric radii ranging from 3 to 16\,kpc, some also placing it near to the Sun (e.g. Dias \& L\'epine~\citealp{dias_lepine_2005}).  Different investigations have likewise inferred a wide range of values of the pattern speed, with estimates ranging from 10 to 30\,\kmskpc\ (e.g.\ Gies \& Helsel~\citealp{gies_helsel_2005}). 
There is not yet a consensus from observations that our Galaxy really does have a global spiral pattern (grand-design spiral arms) rather than just many short arms segments a few kpc long.
Even with the former there is debate whether there are four or two arms. A recent compilation and synthesis of the literature is given by Vall\'ee~\citep{vallee_2008}.

Several studies have implicated passages of the Sun through spiral arms in mass extinctions and/or climate change.  Possible mechanisms are similar to those posited in the case of disk plane crossings, namely encounters with supernovae, gravitational perturbations of the Oort cloud and comet capture.  These studies all assume that there is a grand-design spiral structure, although they differ in their assumptions about the shape, size and speed of the spiral pattern and number of arms. Some turn the problem around and assume that extinctions or ice ages were caused by arm crossings and use this to try and constrain the spiral structure.

\begin{figure}[htb]
\begin{center}
\includegraphics[scale=0.40, angle=0]{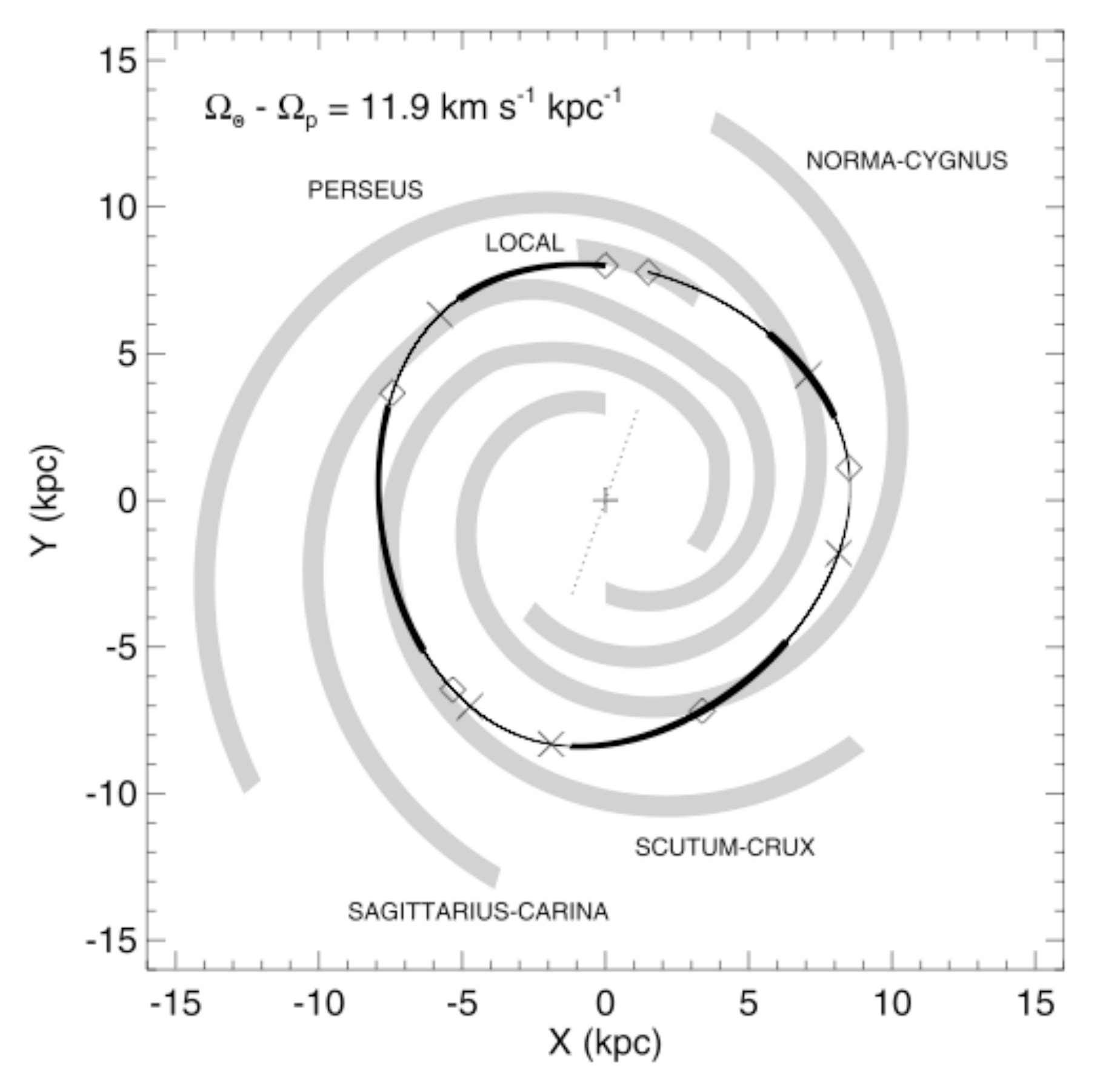}
\caption{Path of the Sun relative to the spiral pattern in the Galaxy as reconstructed by Gies \& Helsel~\citep{gies_helsel_2005} using \omp = 14.4\,\kmskpc. The Galactic centre is at $(0,0$) and the Sun currently at $(0,8.5)$ (the stars and arms rotate clockwise). The diamonds 
mark time intervals of 100\,Myr, crosses selected mass extinctions and the thick solid line icehouse intervals.
Reproduced from Gies \& Helsel~\citep{gies_helsel_2005} with kind permission of D.R.\ Gies and the AAS. 
}  
\label{gh_fig3}
\end{center}
\end{figure}

Gies \& Helsel~\citep{gies_helsel_2005} use the model for the disk gravitational potential from Dehnen \& Binney~\citep{dehnen_binney_1998a} and for the solar motion derived from Hipparcos data by Dehnen \& Binney~\citep{dehnen_binney_1998b} to derive the motion of the Sun over the past 500\,Myr.  Adopting a four-arm model of the spiral structure (with an additional local arm segment) they 
determine when passages occurred for various values for the pattern speed. They compare these with the midpoints of four ice age epochs. (They offer several different times for the occurrence of the ice ages from different sources, with discrepancies of up to tens of Myr.) Using a value for the spiral arm pattern speed of \omp = 20\,\kmskpc\ (in the middle of the range of published estimates), they show that there is little correlation between arm crossings and the ice age midpoints. Testing a range of pattern speeds, as well as values for $R_0$ and the disk scale length in the gravitational potential model of the disk, they obtain a better coincidence when adopting a pattern speed of \omp = 14.4\,\kmskpc (Figure~\ref{gh_fig3}).  (They keep the solar angular motion fixed at $\Omega_{\odot} = 26.3$\,\kmskpc\ in their simulations; it is the relative angular speed of the Sun to the spiral pattern, $\Omega_{\odot} - \Omega_p$, which is relevant.)  This gives arm crossing midpoints at 80, 156, 310 and 446\,Myr BP. However, there are three parameters which can be varied in their model (plus four alternative ice age dating schemes are considered) whereas the quality of the fit is judged based on just four ice age epochs. In other words, the model has considerable degrees of freedom to be constrained by little data, so it should not be hard to obtain a good fit even if the model parameters are constrained in their range. The ice ages are long and the spiral arms wide (they adopt 0.75\,kpc), so some coincidence is almost inevitable. 

Gillman \& Erenler~\citep{gillman_erenler_2007} examine the temporal distribution of numerous geological markers (including extinctions and impact craters) over the past 700\,Myr. By wrapping these at a period of 180\,Myr, they notice that the events fall into three reasonably well-separated ``zones''. If this period is associated with the period between arm passages, then the zones correspond to different phases of the orbit between passages. A fit with a time-domain model (a generalized linear model) gives a period of $175.96 \pm 0.43$\,Myr. (This surprisingly small uncertainty is the formal fitting error as reported: it does not explicitly account for the uncertainties in the geological dating.)  Adopting a four-arm model of the Galaxy with rigid rotation, they identify this figure as the period between arm crossings, which implies \omp\,=\,18.2\,\kmskpc. However, the geological events are not concentrated around arm crossings. As their Figure~2 shows, the events are spread across the whole phase of the purported crossing period, so the relevance of the arms is not obvious. The importance of the period of 176\,Myr seems to be that it divides the geological events into the three zones, but neither the statistical nor astronomical significance of these is clear.  There is no significance analysis of this or any other period, so its uniqueness is hard to assess. A connection with the spiral structure of this model would also require the spiral pattern to have rotated rigidly and not evolved for the past 700\,Myr (four arm crossings).

Shaviv~\citep{shaviv_2003} studied the correlation between cosmic ray fluxes, ice age epochs and spiral structure.  He suggests that there is a causal connection and, depending on the data used, derives a periodicity for spiral arm crossings of between $134 \pm 22$ and $163 \pm 50$\,Myr, with a best fit of the climate proxy data to a spiral arm model yielding $143 \pm 5$\,Myr (using a pattern speed of \omp\,=\,16\,\kmskpc). The fit indicates a lag of the mid-point of glaciations behind arms crossings of $33 \pm 20$\,Myr.  He further concludes that the data indicate the spiral pattern to have been stable over the past billion years, although this assumes the cosmic ray/spiral arm mechanism to be responsible for the climate record (and this seems incompatible with the Myr time lag). This work is based on interpreting the observed clustering in meteorite cosmic ray exposure ages (as measured by the $^{41}$K$/^{40}$K isotope ratio) as evidence for variability in the cosmic ray flux. This is a rather indirect method with a chain of assumptions, and it has been argued that these data are consistent with no clustering and thus no significant cosmic ray variability (Jahnke~\citealp{jahnke_2005}).

The model for the solar motion of Svensmark~\citep{svensmark_2006} discussed in section~\ref{z_motion} is one dimensional, but it uses a potential for the disk which includes two perturbations, considered to be spiral arm crossings.  The increased mass of the spiral arms leads to an acceleration of the vertical motion, with the epoch, duration and amplitude of the perturbation being free parameters in his fit to the Earth temperature proxy data. (Only two perturbations are included based on the prior evidence of the times of spiral arm crossings.)  The inferred times of the arm crossings are 31 and 142\,Myr BP. However,  as I already discussed in section~\ref{z_motion}, because the terrestrial temperature is assumed to depend on distance from the Galactic plane, this does not provide independent evidence that the solar motion (let alone cosmic rays) triggers climate change.

An arm crossing per se does not automatically imply the solar system will experience anything fundamentally different.  Torbett~\citep{torbett_1989} notes that the stellar density inside spiral arms is a factor of only about 1.1 higher than outside the arms. Svensmark~\citep{svensmark_2006} quotes values from the literature of 1.5--1.8 and 1.5--3 for external galaxies. Scoville \& Sanders~\citep{scoville_sanders_1986} estimate the probability of having a ``close'' encounter with a GMC inside a spiral arm as only about 0.1 per arm crossing. If crossings occured on average every 100\,Myr, there is a probability of $0.9^5=0.59$ of no close GMC encounter at all in the past 500\,Myr. Of course, this depends on what one regards as an ``encounter'', and as already discussed the low likelihood, $P(D | H)$, of a hypothesis (without assessing the alternatives) is insufficient to reject it (see section~\ref{ts_fun}).  Leitch \& Vasisht~\citep{leitch_vasisht_1998} estimate the number of supernovae encountered during a spiral arm passage at 0.5, assuming that the supernova only has an effect on the Earth if it passes within 10\,pc (Ellis \& Schramm~\citealp{ellis_schramm_1995}).  This is just an order-of-magnitude figure (not a probability) from a simple volume calculation using the supernova rate (1/30\,yr$^{-1}$ in the whole Galaxy), progenitor lifetime (10\,Myr) and scale height (100\,pc), plus the spiral arm length, width and pattern speed. The true average rate could easily differ from this by an order-of-magnitude or more.  Moreover, GMCs, star formation and supernovae also occur in the disk outside of spiral arms, so even taking the ``star formation region encounter hypothesis'' to be true, we may not expect a very high correlation of geological events with spiral arm transits.

All of these attempts to associate climate change or extinctions with spiral arm crossings are very sensitive to the exact morphology and pattern speed of the spirals arm. Yet the spiral structure of our Galaxy is poorly known: there is still a debate over whether it has a four-arm or two-arm structure, and estimates of the pattern speed vary by a factor of three (12--30\,\kmskpc\ in Table 3 of Shaviv~\citealp{shaviv_2003}; see also Perryman~\citealp{perryman_2009} section 9.7).  Coupled to this is the uncertainty in the corotation radius of the pattern: the nearer it lies to the Sun, the lower the relative velocity of the Sun with respect to the arms and so the lower the frequency of arm crossings.  Moreover, the studies described above assume that the spiral pattern has rotated with a fixed angular speed and with a fixed pattern over hundreds of millions of years.  Yet some N-body simulations predict that spiral arms are unstable, showing significant changes in their structure in less than a rotation period (e.g.\ Sellwood \& Carlberg~\citealp{sellwood_carlberg_1984}).

Another implicit assumption is that the Sun's motion can be described by a smooth gravitational potential and has not experienced any close encounter for hundreds of Myr. There are thousands of known GMCs in the disk which could give the Sun an additional acceleration. The clouds would have now dispersed so it would be almost impossible to reconstruct these events even with accurate stellar kinematics.
Whatever the source, Wielen~\citep{wielen_1977} found that the dispersion velocity of stars increases with time due to orbital diffusion. This produces a velocity change of order 10\,\kms\ in a single orbit of the Sun around the Galaxy.

In summary, it is likely that the Sun has crossed the spiral arms up to a handful of times over the Phanerozoic.  But the specific conclusions of the cited studies of a connection to climate change or mass extinctions are very sensitive to the very uncertain structure, kinematics and evolution of the spiral arms, and to the uncertainties in the Galactic potential used to reconstruct the motion of the Sun. 
It is premature to draw a connection, let alone use any apparent one to provide support for a specific extinction/climate change mechanism such as cosmic rays or Oort cloud perturbation.  But we can draw at least one useful conclusion from these studies: As the arms are unlikely to show a perfectly symmetric structure and fixed rotation relative to the Sun, it is unlikely that arm crossings are periodic, and almost certainly not frequent enough to explain the 26 or 62\,Myr periods (or values near these) discussed in section~\ref{phan_periods}. Moreover, using these studies we can identify what better astronomical data we need to make progress.  Some studies attempted to use the geological record to formally fit (or loosely constrain) models for the spiral structure. In my opinion, given the uncertainties in the data and the large degree of freedom in the models (or the use of numerous assumptions), this is, at best, inconclusive. Although it is frustrating that the astronomical data are not yet adequate, it is nonetheless essential that independent astronomical data be used to model astronomical phenomena, which only then are compared with the geological data.

\subsection{Radial variations}

Goncharov \& Orlov~\citep{goncharov_orlov_2003} claim that 13 mass extinction events have a ``repetition interval'' of $183 \pm 3$\,Myr.
It is not clear what they mean by this, because the events are not periodic with this period (their extinction dates range from 2--469\,Myr BP, much less than $12 \times 183$\,Myr) and the events (their Figure 1) do not cluster around this period.  They suggest that extinctions could be described by the radial motion of the Sun in its orbit around the Galactic centre: they quote a model in which $R_0$ varies from 7.96 to 8.15\,kpc with a period of 183\,Myr. However, a superposition of the variation of $R_0$ over time in this model with these extinction events (their Figure 3) shows only a very slight clustering of events at the apocentre and pericentre positions. They note that if only a subset of the events are retained then the clustering is better. But if one arbitrarily select events then I suspect it would be possible to make a number of periods ``fit'' the data. Without a systematic analysis of the significance of the clustering at a range of periods and the sensitivity to the data retained, it is not possible to draw any firm conclusions from this work. Of course, it is {\em a priori} possible that a subset of extinction events do show periods and are caused by a periodic mechanism. But unless one can find a signature in the data to decide which events to select {\em independently} of the period of the mechanism, the data themselves cannot be taken to support this hypothesis. The hypothesis ``some of the events are periodic'' is simply too general and too flexible to have much predictive power and thus to get much support from the data.

%%%%%%%%%%%%%%%%%%%%%%%%%%%%%%%%%%%%%%%%%%%%%%%%%%%%
\section{Improving the situation}\label{improving}

In section~\ref{reliability} I discussed various sources of uncertainty in the data and their calibration. Improvements in the dating of mass extinctions and craters and in the completeness and temporal resolution of the fossil record would certainly make the studies described more conclusive. Removing sources of bias and better calibrating proxies to ensure that they measure what we want them to measure are also important. These may improve with time, but at some level nature sets fundamental limits. On the data analysis side, I discussed in section~\ref{ts_fun} issues in time series analysis and hypothesis testing. Some of the techniques used have been demonstrated to give spurious results and other techniques should be (more) rigorously tested. Moreover, I have argued that better astronomical models and better data to fit them are required. What prospects are there for improving the situation?

We need to determine two things in particular more accurately: (1) the path of the Sun through the Galaxy, (2) the structure, velocity and evolution of the spiral arms. The first requires a much better determination of the Galactic potential and how it may have evolved, as well as a more accurate determination of the current phase space coordinates of the Sun (position, velocity). The second requires that we better trace the position and velocity of the spiral arms. Combined with a knowledge of the Galactic potential we can (to some degree) wind the arms back over the past half Gyr.

These points will be addressed by the upcoming ESA mission Gaia (Turon et al.~\citealp{turon_etal_2005},
Lindegren et al.~\citealp{lindegren_etal_2008}, 
 Bailer-Jones~\citealp{cbj_2009}).\footnote{\url{http://www.rssd.esa.int/Gaia}} Due for launch in 2012, Gaia will measure accurate positions, distances and proper motions for essentially all objects in the sky brighter than magnitude G=20 (G is the broad 350--1000\,nm observation band), some $10^9$ stars and a million or so galaxies and quasars.  The parallax accuracy is 12--25\,\uas\ at G=15 and 100--300\,\uas\ at G=20.  (These are also the approximate proper motion accuracy in \uas/year.)  This translates to distances accurate to 1\% for 11 million stars out to 800\,pc, or accurate to 10\% for 150 million stars out to 8\,kpc. (This compares to just 200 stars currently which have parallaxes measured to better than 1\%, all of which are within 10\,pc.)  Gaia also measures radial velocities to a few \kms\ for objects down to G=17. From the onboard low resolution photometry we can estimate stellar parameters (effective temperature, surface gravity, metallicity, line-of-sight extinction) from which kinematic and chemical tracers can be selected.  
Using K giants we can determine the gravitational potential of the Galactic disk out to several kpc in galactocentric radius: For a K giant at 6 kpc (G=15) we can measure its distance to an accuracy of 2\% and its velocity to 1\,\kms. As this is a kinematical measurement of the potential, it includes the dark matter component. 

Through its mapping of the Galaxy, Gaia will improve the determinations of the distance to the Galactic centre and the velocity of the Sun. This will better constrain models of the solar motion (e.g.\ the epicyclic model) and so determine more accurately the amplitude and period of the motion about the disk and in the radial direction and the likely departures from pure periodic motion.

We will also be able to use Gaia data to measure the position and velocities of the spiral arms themselves from  observations of their OB star population, without assuming a rotation curve or needing to know the interstellar extinction.  For an OB star 5\,kpc from the Sun observed through 4 magnitudes of extinction, Gaia will determine its distance to an accuracy of 13\% and its space velocity to 1\,\kms.  Gaia can do this for some 50\,000 OB stars within a few kpc. Gaia will also trace thousands of open clusters and star forming regions in the disk, for which more accurate ensemble distance and velocity estimates are possible.

Quantification of the improvements which Gaia will lead to must await future studies. But there is no doubt that when the data arrive, many of the studies described above can be improved upon and the conclusions reassessed.

%%%%%%%%%%%%%%%%%%%%%%%%%%%%%%%%%%%%%%%%%%%%%%%%%%%%
\section{Conclusions}\label{conclusions}

I have examined the evidence for an astronomical role in biological evolution and climate over the Phanerozoic eon (past 545\,Myr). The objective was to examine the plausible mechanisms for change, their possible astronomical root causes in the motion of the Sun and Earth and whether they are supported by the geological record. Based on this, I draw the following conclusions.
\vspace*{-0.5em}
\begin{itemize}
\item There is no good evidence for a periodicity in the biodiversity, extinction or cratering record with a period in the range 25--33\,Myr. Most studies which have claimed such periods have been affected by issues of data selection, dating errors, methodological flaws, lack of adequate significance/hypothesis testing, or a combination of these.
\item There is reasonable evidence supporting a $62 \pm 3$\,Myr periodicity in the biodiversity data for the period 520--150\,Myr\,BP, although its robustness to dating, calibration and selection errors needs to be explored further.  There is some question of whether this signal is measuring variations in biodiversity rather than in fossil preservation.  The period is similar in magnitude to the $z$-oscillation period of the Sun about the Galactic plane (although this is not known precisely).  It has been suggested that cosmic rays from a Galactic bow shock due to the motion of the Galaxy may be a cause of extinctions (perhaps via climate change). However, there is not yet any evidence for this mechanism nor for a 62\,Myr periodic variation in the cosmic ray flux reaching the Earth, and the lack of periodicity in biodiversity in the last 150\,Myr remains unexplained.
\item There is no evidence that either Galactic plane crossings, spiral arm crossings or any other aspect of the solar motion play a significant role in climate change or mass extinctions, whether by cosmic rays, supernovae, impacts from Oort cloud perturbation or any other mechanism.  This holds whether the spiral arm crossings are periodic or not.  
This conclusion is a consequence of the significant uncertainties and assumptions in the astronomical mechanisms, as well as uncertainties in interpreting the geological records. 
The claims for a spiral arm connection are particularly sensitive to their poorly known structure, kinematics and evolution.
Either these mechanisms are not relevant, or the spiral arms/Galactic plane do not provide a fundamentally different environment for the solar system. To make progress on this, it is imperative that the solar motion is derived independently of the geological record.
\item There is no direct evidence that any of the discussed extraterrestrial (or terrestrial) mechanisms have had a periodic influence on climate or biodiveristy with periods of a Myr or longer. The only good evidence we have for a recurring astronomical influence on climate is from ice cores and foraminifera fossils, which suggest that variations in the Earth's orbit (eccentricity, obliquity, precession) over the past 3\,Myr have influenced global temperatures and ice ages with periods of tens of kyr.
\item Statistically, sufficiently nearby supernova blasts or gamma-ray bursts could have occured a few times during the Phanerozoic.  While studies have shown that {\em could} cause widespread extinction, there is very little evidence that they actually have.
\item Due in part to methodological problems with the studies, there is no good evidence that cosmic rays have a significant influence on the Earth's climate on either Myr or decadal timescales. (They do not explain the majority of post-industrial global warming.)  There is nonetheless the indication that cosmic rays could have some impact, at least on 1--10\,yr timescales. But the mechanism of cloud nucleation via cosmic rays remains sketchy and undemonstrated. Other possible effects of cosmic rays need to be explored.
\item Both mass volcanism and large asteroid/comet impacts have occurred in the past and these have probably caused widespread extinction and maybe short-term (10--$10^5$\,yr) climate change.  The K-T extinction 65\,Myr BP was almost certainly caused in part by a a large impact, although volcanism may have contributed too. There is no good geological evidence (iridium, tektites, shocked quartz and a crater) for impacts having caused other extinctions.  There is no evidence for a periodicity in impacts and so no need to invoke a periodic Oort cloud perturbation mechanism.  Volcanism coincides very closely with several mass extinctions, but there is limited direct evidence for a causal connection.
Changes in climate, sea level and sea oxygen levels as a result of plate tectonics have probably played a role in evolution and show temporal coincidence with many mass extinction.
\item As alternative hypotheses, there is evidence suggesting that some of the variability in the fossil record may reflect variations in the efficiency of fossilization rather than variations in biodiversity itself.  Purely biological models can explain rare mass extinctions as just the tail of a distribution of extinctions resulting from population dynamics.  Other models show how apparent periods can be a signature of ecosystem response or a result of the internal dynamics of multi-species evolution, without needing to invoke an external driver.
\end{itemize}

Much of this work depends on inference from sparse or noisy data. This is a complex procedure, with many difficult choices to make. Some issues which arose are fundamental to the data analysis procedure and are now summarized.
\vspace*{-0.5em}
\begin{itemize}
\item Dating errors, date rounding and sample contamination can generate spurious periods when using some time series methods.
\item The conclusions of some studies depended on having made somewhat arbitrary selections or corrections of data.
\item Some non-periodic but non-random models (e.g.\ a moving average) can produce an apparent period when analysed by some time series techniques.
\item Orthodox hypothesis testing calculates a $p$-value, the probability of observing some statistic given some (null) hypothesis ($H_0$). Oddly, this often depends on the probability of observing some unseen data, e.g.\ the probability of reaching a power {\em or more} in a periodogram. (More oddly, orthodox hypothesis testing never explicity tests the hypothesis of interest.)
Even if we equate the $p$-value to the probability of getting the data given some null hypothesis, $P(D|H_0)$, this is not the probability of the null hypothesis given the data, $P(H_0|D)$, which is the quantity we are interested in. A low value of $P(D|H_0)$ does not rule out $H_0$.
\item Even if we reject some null hypothesis, this does not mean some alternative is true.  There may be other untested hypotheses which are better supported by the data.  There are many plausible null hypotheses, or ``random'' data sets, for time series data, so rejecting just one or two of these does not make the periodic model true (Figure~\ref{hypotest}).  Determination of a low value of $P(D|H_0)$ is useful, but it is insufficient (``incomplete inference'').
\item 
We can only rule out a hypothesis if we can show that an alternative is more plausible. 
The only reliable way to test a hypothesis is to compare it with other hypotheses, that is, compare their $P(D|H)$ values. It is sometimes difficult to specify all plausible hypotheses (which is often why one resorts to orthodox testing), but if we can, we can do a full Bayesian calculation to calculate $P(H|D)$ (equation~\ref{eqn:phd}). This also allows us to explicitly accommodate different priors on the hypotheses.
\end{itemize}

The search for periodicities which point to a single cause of extinction or climate change is luring.  But is there good reason to expect a single, universal cause?  There are many processes which could contribute to changes in biodiversity or climate and there is evidence that some of these actually have.  Perhaps a coincidence of processes is necessary to cause the biggest extinctions. All of these geological, astronomical and biological processes have a size distribution.  Given the limited sensitivity of proxies and the fossil record, it may instead be that we only observe large events which stand out above the background.  There are many reasons why we do not expect these mechanisms to give rise to periodicities in the geological data, and it now seems that evidence for periodicities is indeed lacking. Even possible astronomical mechanisms -- which involve numerous assumptions -- are a priori unlikely to be strictly periodic. While astronomical mechanisms may have triggered some mass extinctions and climate change, there is little to support them as a universal or even significant cause.

%%%%%%%%%%%%%%%%%%%%%%%%%%%%%%%%%%%%%%%%%%%%%%%%%%%%
\subsubsection*{Acknowledgements}

I would like to thank Georg Feulner, David Hogg, Knud Jahnke, Barrie Jones, Rainer Klement, Adrian Melott, Bill Napier and Stephen Stigler for useful comments on a draft of this manuscript.

%%%%%%%%%%%%%%%%%%%%%%%%%%%%%%%%%%%%%%%%%%%%%%%%%%%%


\begin{thebibliography}{99}

%\begin{flushleft}
%\begin{verse}

\bibitem[2002]{abbott_isley_2002} 
Abbott D.H., Isley A.E., 2002, {\em Extraterrestrial influences on mantle plume activity}, Earth and Planetary Science Letters 205, 53--62

\bibitem[2008]{alroy_2008}
Alroy J., 2008, {\em Dynamics of origination and extinction in the marine fossil record}, Proceedings of the National Academy of Sciences 105, 11536--11542

\bibitem[2003]{alvarez_2003}
Alvarez W., 2003, {\em Comparing the evidence relevant to impact and flood basalt at times of major mass extinctions}, Astrobiology 3, 153--161

\bibitem[1984]{alvarez_muller_1984}
Alvarez W., Muller R.A., 1984, {\em Evidence from crater ages for periodic impacts on the earth}, Nature 308, 718--720

\bibitem[1980]{alvarez_etal_1980}
Alvarez L.W., Alvarez W., Asaro F., Michel H.V., 1980, {\em Extraterrestrial cause for the Cretaceous-Tertiary extinction}, Science 208, 1095--1108

\bibitem[2008]{arens_west_2008}
Arens N.C., West I.D., 2008, {\em Press--pulse: a general theory of mass extinction?}, Paleobiology 34, 456--471

\bibitem[2009]{cbj_2009}
Bailer-Jones C.A.L., 2009, {\em What will Gaia tell us about the Galactic disk?}, in The Galaxy Disk in Cosmological Context, Proceedings of the International Astronomical Union, IAU Symposium, Volume 254, Edited by J. Andersen, J. Bland-Hawthorn, B. Nordstr\"om, CUP, pp. 475--482

\bibitem[1985]{bahcall_bahcall_1985} 
Bahcall J.N., Bahcall S., 1985, {\em The Sun's motion perpendicular to the galactic plane}, Nature 316, 706--708

\bibitem[2006]{bambach_2006} 
Bambach R.K., 2006, {\em Phanerozoic biodiversity mass extinctions}, Ann. Rev. Earth Planet. Science 34, 127--155

\bibitem[2003]{berger_2003}
Berger J.O., 2003, {\em Could Fisher, Jeffreys and Neyman have agreed on testing?}, Statistical Science 18, 1--32

\bibitem[2002]{carslaw_etal_2002}
Carslaw K.S., Harrison R.G., Kirkby J., 2002, {\em Comsic rays, clouds, and climate}, Science 298, 1732--1737

\bibitem[1995]{cincotta_etal_1995}
Cincotta P.M., M\'endez M., N\'u\~nez J.A., 1995, {\em Astronomical time series analysis 1. A search for periodicity using information entropy}, ApJ 449, 231--235

\bibitem[2004]{chapman_2004}
Chapman C.R., 2004, {\em The hazard of near-Earth asteroid impacts on earth}, Earth and Planetary Science Letters 222, 1--15

\bibitem[2005]{christensen_2005}
Christensen R., 2005, {\em Testing Fisher, Neyman, Pearson and Bayes}, The American Statistician 59, 121--126

\bibitem[1982a]{clube_napier_1982a} 
Clube S.V.M., Napier W.M., 1982a, {\em Spiral arms, comets and terrestrial catastrophism}, Quarterly Journal of the Royal astronomical Society 23, 45--66

\bibitem[1982b]{clube_napier_1982b} 
Clube S.V.M., Napier W.M., 1982b, {\em The role of episodic bombardment in geophysics}, Earth and Planetary Science Letters 57, 251--262

\bibitem[2007]{cornette_2007}
Cornette, J.L., 2007, {\em Gauss--Van\'{\i}\v{c}ek and Fourier transform spectral analyses of marine diversity}, Computing in Science and Engineering, July/August 2007 61--63

\bibitem[2008]{dali_zi_2008}
Da-li K., Zi Z., 2008, {\em A study of the scale height of the thin Galactic disk in the solar neighbourhood}, Chinese Astronomy and Astrophysics 32, 360--368

\bibitem[2004]{damon_laut_2004}
Damon P.E., Laut P., 2004, {\em Pattern of strange errors plagues solar activity and terrestrial climate data}, Eos 85, 370--374

\bibitem[1964]{dansgaard_1964}
Dansgaard W., 1964, {\em Stable isotopes in precipitation}, Tellus 16, 436--468

\bibitem[1984]{davis_etal_1984}
Davis M., Hut P., Muller R.A., 1984, {\em Extinction of species by periodic comet showers}, Nature 308, 715--717

\bibitem[2005]{dias_lepine_2005}
Dias W.S., L\'epine J.R.D., 2005, {\em Direct determination of the spiral pattern rotation speed of the Galaxy}, ApJ 629, 825--831

\bibitem[1998a]{dehnen_binney_1998a}
Dehnen W., Binney J.J., {\em Mass models of the Milky Way}, MNRAS 294, 429--438

\bibitem[1998b]{dehnen_binney_1998b}
Dehnen W., Binney J.J., {\em Local stellar kinematics from Hipparcos data}, MNRAS 298, 387--394

\bibitem[1995]{ellis_schramm_1995}
Ellis J., Schramm D.N., 1995, {\em Could a nearby supernova explosion have caused a mass extinction?}, Proc.\ Natl.\ Acad.\ Sci.\ USA 92, 235--238 

\bibitem[1961]{epstein_etal_1961}
Epstein S., Buchsbaum R., Lowenstam H., Urey H.C., 1961, {\em Carbonate--water isotopic temperature scale}, Bulletin of the Geological Society of America 62, 417--426

\bibitem[2009]{erlykin_eta_2009}
Erlykin A.D., Sloan T., Wolfendale A.W., 2009, {\em Solar activity and the mean global temperature}, 
Environ. Res. Lett. 4, 014006

\bibitem[2003]{erwin_2003}
Erwin D.H., 2003, {\em Impact at the Permo-Triassic boundary: a critical evaluation}, Astrobiology 3, 67--74

\bibitem[2009]{feulner_2009} 
Feulner G., 2009, {\em Climate modelling of mass-extinction events : a review}, International Journal of Astrobiology, in press (this volume)

\bibitem[1925]{fisher_1925}
Fisher R.A, 1925, {\em Statistical methods for research workers}, Oliver \& Boyd, Edinburgh

\bibitem[1991]{friis_lassen_1991}
Friis-Christensen E., \& Lassen K., 1991, {\em Length of the solar cycle: An indicator of solar activity closely associated with climate}, Science 254, 698--700

\bibitem[2009]{fuchs_etal_2009} 
Fuchs B., Dettbarn C., Rix H. -W., Beers T. C., Bizyaev D., Brewington H., Jahreiss H., Klement R., Malanushenko E., Malanushenko V., Oravetz D., Pan K., Simmons A., Snedden S.,
2009, {\em The kinematics of late type stars in the solar cylinder studied with SDSS data}, AJ 137, 4149--4159

\bibitem[2001]{garwin_charpak_2001}
Garwin R.L., Charpak G., 2001, {\em Megawatts and Megatons}, New York: Alfred A.\ Knopf

\bibitem[2005]{gies_helsel_2005}
Gies D.R., Helsel J.W., 2005, {\em Ice age epochs and the Sun's path through the Galaxy}, ApJ 626, 844--848

\bibitem[2007]{gillman_erenler_2007} 
Gillman M., Erenler H., 2007, {\em The galactic cycle of extinction}, International Journal of Astrobiology 7, 17--26

\bibitem[2003]{glikson_2003}
Glikson A., 2003, Comment on Abbott \& Isley (2002), Earth and Planetary Science Letters 215, 425--427

\bibitem[1994]{glen_1994}
Glen W., (editor), 1994, {\em The mass-extinction debates: How science works in a crisis}, Stanford University Press

\bibitem[2003]{goncharov_orlov_2003} 
Goncharov G.N., Orlov V.V., 2003, {\em Global repeating events in the history of the Earth and the motion of the Sun in the Galaxy}, Astronomy reports 47, 925--933

\bibitem[2005]{ics_2005}
Gradstein F., Ogg J., Smith A., 2005, {\em A geologic time scale 2004}, Cambridge University Press

\bibitem[2005]{gregory_2005}
Gregory, P., 2005, {\em Bayesian logical data analysis for the physical sciences}, Cambridge University Press

\bibitem[1992]{gregory_loredo_1992}
Gregory P.C., Loredo T.J., 1992, {\em A new method for the detection of a periodic signal of unknown shape and period}, ApJ 398, 146--168

\bibitem[1985]{grieve_etal_1985}
Grieve R.A.F., Sharpton V.L., Goodacre A.K., Garvin J.B., 1985, {\em A perspective on the evidence for periodic cometary impacts on Earth}, Earth and Planetary Science Letters 76, 1--9

\bibitem[1988]{grieve_etal_1988} 
Grieve R.A.F., Sharpton V.L., Rupert J.D., Goodacre A.K., 1988, {\em Detecting a periodic signal in the terrestrial cratering record}, Proceedings of the $18^{th}$ LPSC, 375--382

\bibitem[1989]{hallam_1989}
Hallam A., 1989, {\em The case for sea-level change as a dominant causal factor in mass extinction of marine invertebrates}, Phil.\ Tran.\ Royal Society Series B 325, 437--455

\bibitem[2004]{hallam_2004}
Hallam A., 2004, {\em Catastrophes and lesser calamaties. The causes of mass extinctions}, Oxford University Press

\bibitem[2008]{harris_2008}
Harris A., 2008, {\em What Spaceguard did}, Nature 453, 1178--1179

\bibitem[1976]{hays_etal_1976}
Haye J.D., Imbrie J., Shackleton N.J., 1976, {\em Variations in the Earth's orbit: Pacemaker of the ice ages}, Science 194, 1121--1132

\bibitem[1989]{heisler_tremaine_1989}
Heisler J., Tremaine S., 1989, {\em How dating uncertainties affect the detection of periodicity in extinctions and craters}, Icarus 77, 213--219

\bibitem[1991]{hildebrand_etal_1991}
Hildebrand A. R., Penfield G.T., Kring D.A., Pilkington M., Camargo Z.A., Jacobsen S.B., Boynton,W.V., 1991, {\em Chicxulub Crater: A possible Cretaceous/Tertiary boundary impact crater on the Yucatan Penninsula, Mexico}, Geology 19, 867--871

\bibitem[1939]{hoyle_lyttleton_1939}
Hoyle F., Lyttleton R.A., 1939, {\em The effect of interstellar matter on climatic variation}, Proc.\ Cambridge Philosophical Society 35, 401--415

\bibitem[1978]{hoyle_1978}
Hoyle F., Wickramasinghe C., 1978, {\em Comets, ice ages, and ecological catastrophes}, 
Astrophysics and Space Science 53, 523--526

\bibitem[1984]{hut_1984}
Hut P., 1984, {\em How stable is an astronomical clock that can trigger mass extinctions on Earth?}, Nature 311, 638--641

\bibitem[2007]{ipcc_2007}
IPCC, 2007, {\em Climate change 2007: The physical science basis. Contribution of working group I to the fourth assessment report of the intergovernmental panel on climate change}, Solomon et al. {eds}, CUP

\bibitem[2005]{jahnke_2005}
Jahnke K., 2005, {\em On the periodic clustering of cosmic ray exposure ages of iron meteorites},
unpublished manuscript, astro-ph/0504055v1 

\bibitem[2003]{jaynes_2003}
Jaynes E.T., 2003, {\em Probability theory. The logic of science}, Cambridge University Press

\bibitem[2000]{jetsu_pelt_2000} 
Jetsu L., Pelt J., 2000, {\em Spurious periods in the terrestrial impact crater record}, A\&A 353, 409--418

\bibitem[2000]{jorgensen_hansen_2000}
J{\o}rgensen T.S., Hansen A.W., 2000, 
Comments on Svensmark \& Friis-Christensen~\citep{svensmark_friis_1997},  J.\ Atmospheric and Solar--Terrestrial Physics 62, 73--77

\bibitem[1991]{kauffman_johnsen_1991}
Kauffman S.A., Johnsen S., 1991, {\em Coevolution to the edge of chaos: Coupled fitness landscapes, poised states, and coevolutionary avalanches}, J.\ theoretical Biology 149, 467--505

\bibitem[2007]{kirkby_2007}
Kirkby J., 2007, {\em Cosmic rays and climate}, Surveys in Geophysics 28, 333-375

\bibitem[1984]{kitchell_pena_1984} 
Kitchell J.A., Pena D., 1984, {\em Periodicity of extinctions in the geologic past: Deterministic versus stochastic explanations}, Science 226, 689--692

\bibitem[2002]{kristjansson_etal_2002} 
Kristj\'ansson J.E., Staple A., Kristiansen J., Kaas E., 2002, {\em A new look at possible connections between solar activity, clouds and climate}, Geophysical Research Letters 29, 221--224

\bibitem[2003]{laut_2003}
Laut P., 2003, {\em Solar activity and terrestrial climate: an analysis of some purported correlations}, Journal of atmospheric and solar--terrestrial physics 65, 801--812

\bibitem[2005]{lisiecki_raymo_2005}
Lisiecki L.E., Raymo M.E., 2003, {\em A Pliocene--Pleistocene stack of 57 globally distributed benthic \oproxy\ records}, Paleoceanography 20, PA1003

\bibitem[1964]{lin_shu_1964}
Lin C.C., Shu F.H., 1964, {\em On the spiral structure of disk galaxies}, ApJ 140, 464--655

\bibitem[1938]{lindblad_1938}
Lindblad B., 1938, {\em On the theory of spiral structure in the nebulae}, Zeitschrift f\"ur Astrophysik 15, 124--136

\bibitem[2008]{lindegren_etal_2008}
Lindegren L., Babusiaux C., Bailer-Jones C.A.L. Bastian U.,  
Brown A.G.A., Cropper M., Hog E., Jordi C., Katz D., van Leeuwen F., Luri X., Mignard F., de Bruijne J., Prusti, T., 2008, {\em The Gaia mission: science, organization and present status}, in {\em A Giant Step: from Milli- to Micro-arcsecond Astrometry}, Proceedings of the International Astronomical Union, IAU Symposium, W.J.\ Jin, I.\ Platais, M.A.C.\ Perryman (eds.), Volume 248, 217--223

\bibitem[1998]{leitch_vasisht_1998}
Leitch E.M., Vasisht G., 1998, {\em Mass extinctions and the sun's encounters with spiral arms}, New Astronomy 3, 51--56

\bibitem[2007]{lieberman_melott_2007} 
Lieberman B.S., Melott A.L., 2007, {\em Considering the case for biodiversity cycles: re-examining the evidence for periodicity in the fossil record}, PLoS ONE 8, e759

\bibitem[2009]{lieberman_melott_2009} 
Lieberman B.S., Melott A.L., 2009, {\em Whilst this planet has gone cycling on: what role for periodic astronomical phenomena in large scale patterns in the history of life?}, arXiv:0901.3173

\bibitem[2005]{lockwood_2005} 
Lockwood M., 2005, {\em Solar outputs, their variations and their effects on Earth}, in {\em The Sun, solar analogs and the climate}, Haigh J.D., Lockwood M., Giampapa M.S., Springer, Berlin

\bibitem[2007]{lockwood_froehlich_2007} 
Lockwood M., Fr\"ohlich C., 2007, {\em Recent oppositely directed trends in solar climate forcings and the global mean surface air temperature}, Proc. Royal Society A, 463, 2447--2460

\bibitem[1976]{lomb_1976}
Lomb N.R., 1976, {\em Least-squares frequency analysis of unequally spaced data}, Astrophysics and Space Science 39, 447--462

\bibitem[1985]{lutz_1985}
Lutz T.M., 1985, {\em The magnetic reversal record is not periodic}, Nature 317, 404--407

\bibitem[2000]{marsh_svensmark_2000}
Marsh N.D., Svensmark H., 2000, {\em Low cloud properties influenced by cosmic rays}, Physical Review Letters 85, 5004--5007

\bibitem[1996]{matsumoto_kubotani_1996}
Matsumoto M., Kubotani H., 1996, {\em A statistical test for correlation between crater formation rate and mass extinctions}, MNRAS 282, 1407--1412

\bibitem[1975]{mccrea_1975}
McCrea W.H., 1975, {\em Ice ages and the Galaxy}, Nature 255, 607--609

\bibitem[2007]{medvedev_melott_2007}
Medvedev M.V., Melott A.L., 2007, {\em Do extragalactic cosmic rays induce cycles in fossil diversity?},
ApJ 664, 879--889

\bibitem[2008]{melott_2008} 
Melott A.L., 2008, {\em Long-term cycles in the history of life: Periodic biodiversity in the Paleobiology database}, PLoS ONE 3, e4044

\bibitem[2004]{melott_etal_2004} 
Melott A.L., Lieberman B.S., Laird C.M., Martin L.D., Medvedev M.V., Thomas B.C., Cannizzo J.K., Gehrels N., 
Jackman, C.H., 2004, {\em Did a gamma-ray burst initiate the late Ordovician mass extinction?},
International Journal of Astrobiology 3, 55--61

\bibitem[2003]{morrison_2003}
Morrison D., 2003, {\em Impacts and evolution: Future prospects}, Astrobiology 3, 193--205

\bibitem[2000]{muller_macdonald_2000}
Muller R.A., MacDonald G.J., 2000, {\em Ice ages and astronomical causes}, Springer--Praxis: Chichester

\bibitem[1986]{muller_morris_1986}
Muller R.A., Morris D.E., 1986, {\em Geomagnetic reversals from impacts on the Earth}, Geophysical Research Letters 13, 1177--1180

\bibitem[1988]{napier_1998}
Napier W.N., 1988, {\em NEOs and impacts: The Galactic connection}, Celestial Mechanics and Dynamical Astronomy 69, 59--75

\bibitem[1979]{napier_clube_1979}
Napier W.M., Clube S.V.M., 1979, {\em A theory of terrestrial catastrophism}, Nature 282, 455--459

\bibitem[1983]{negi_tiwari_1983}
Negi J.G., Tiwari R.K., 1983, {\em Matching long term periodicities of geomagnetic reversals and Galactic motions of the solar system}, Geophysical Research Letters 10, 713--716

\bibitem[1997]{newman_1997}
Newman M.E.J., 1997, {\em A model of mass extinction}, J.\ theoretical Biology 189, 235--252

\bibitem[2006]{omerbashich_2006}
Omerbashich M., 2006, {\em A Gauss--Van\'{\i}\v{c}ek spectral analysis of the Sepkoski compendium: no new life cycles}, Computing in Science and Engineering, July/August 2006, pp.\ 26--30 (Erratum in July/August 2007, pp.\ 4--6)

\bibitem[2003]{palle_butler_2002}
Pall\'e E., Butler C.J., 2002, {\em The proposed connection between clouds and cosmic rays: cloud behaviour during the past 50--120 years}, J.\ Atmospheric and Solar--Terrestrial Physics 64, 327--337

\bibitem[1987]{pandey_negi_1987}
Pandey O.P., Negi J.G., 1987, {\em Global volcanism, biological mass extinctions and the galactic vertical motion of the solar system}, Geophys. J.\ Royal astronomical Society 89, 857--867

\bibitem[1987]{patterson_smith_1987} 
Patterson C., Smith C., 1987, {\em Is the periodicity of extinctions a taxonomic artefact?}, Nature 330, 248--252

\bibitem[2009]{perryman_2009}
Perryman M.A.C., 2009, {\em Astronomical applications of astrometry}, Cambridge University Press

\bibitem[2002]{peters_foote_2002}
Peters S.E., Foote M., 2002, {\em Determinants of extinction in the fossil record}, Nature 416, 420--424

\bibitem[1999]{petit_etal_1999}
Petit J.R., J. Jouzel, D. Raynaud, N.I. Barkov, J.-M. Barnola,
2I. Basile, M. Bender, J. Chappellaz, M. Davis, G. Delayque,
M. Delmotte, V.M. Kotlyakov, M. Legrand, V.Y. Lipenkov, C. Lorius,
L. Pepin, C. Ritz, E. Saltzman, and M. Stievenard, 1999, {\em 
Climate and atmospheric history of the past 420,000 years from the
Vostok ice core, Antarctica}, Nature 399, 429--436.

\bibitem[1993]{plotnick_mckinney_1993}
Plotnick R.E., McKinney M.L., 1993, {\em Ecosystem organization and extinction dynamics}, Palaios 8, 202--212

\bibitem[1998]{rampino_1998}
Rampino M.R., 1998, {\em The Galactic theory of mass extinctions: An update}, Celestial Mechanics and Dynamical Astronomy 69, 49--58

\bibitem[1992]{rampino_caldeira_1992}
Rampino M.R., Caldeira K., 1992, {\em Episodes of terrestrial geologic activity during past 260 million years: A quatitative approach}, Celestial Mechanics and Dynamical Astronomy 54, 143--159

\bibitem[1984a]{rampino_stothers_1984a}
Rampino M.R., Stothers R.B., 1984a, {\em Terrestrial mass extinctions, cometary impacts and the sun's motion perpendicular to the galactic plane}, Nature 308, 709--712

\bibitem[1984b]{rampino_stothers_1984b}
Rampino M.R., Stothers R.B., 1984b, Reply to Stigler (1985), Nature 313, 159--160

\bibitem[1985a]{raup_1985a}
Raup D.M., 1985a, {\em Magnetic reversals and mass extinctions}, Nature 314, 341--343

\bibitem[1985b]{raup_1985b}
Raup D.M., 1985b, {\em Rise and fall of periodicity}, Nature 317, 384--385

\bibitem[1984]{raup_sepkoski_1984}
Raup D.M., Sepkoski J.S., 1984, {\em Periodicity of extinctions in the geologic past}, Proc.\ Natl.\ Acad.\ Sci.\ USA 81, 801--805

\bibitem[1986]{raup_sepkoski_1986}
Raup D.M., Sepkoski J.S., 1986, {\em Periodic extinctions of families and genera}, Science 231, 833--836

\bibitem[1988]{raup_sepkoski_1988}
Raup D.M., Sepkoski J.S., 1988, {\em Testing for periodicity of extinction}, Science 241, 94--96

\bibitem[2004]{rahmstorf_etal_2004} 
Rahmstorf S., Archer D., Ebel D.S., Eugster O., Jouzel J., Mauran D., Neu U., Schmidt G.A., Severinghaus J., Weaver A.J., Zachos J., 2004, {\em Cosmic rays, carbon dioxide, and climate}, Eos 85, 38--41

\bibitem[2005]{rohde_muller_2005}
Rohde R.A., Muller R.A., 2005, {\em Cycles in fossil diversity}, Nature 434, 208--210

\bibitem[2004]{royer_etal_2004}
Royer D.L., Berner R.A., Monta\~nez I.P., Tabor N.J., Beering D.J., 2004, {\em CO$_2$ as a primary driver of Phanerozoic climate}, GSA Today 14, 4--10

\bibitem[1974]{ruderman_1974}
Ruderman M.A., 1974, {\em Possible consequences of nearby supernova explosions for atmospheric ozone and terrestrial life}, Science 184, 1079--1081

\bibitem[1982]{scargle_1982}
Scargle J.D., 1982, {\em Studies in astronomical time series analysis 2. Statistical aspects of spectral analysis of unevenly spaced data}, ApJ 263, 835--853

\bibitem[2008]{schaefer_2008}
Schaefer B.E., 2008, {\em A problem with the clustering of recent measures of the distance to the Large Magellanic Cloud}, AJ 135, 112--199

\bibitem[1986]{scoville_sanders_1986}
Scoville N.Z., Sanders D.B., 1986, {\em Observational constraints on the interaction of giant molecular clouds with the solar system}, In Smoluchowski et al.\ (eds), {\em The Galaxy and the Solar System}, Univ.\ Arizona Press, Tucson, pp. 69--82

\bibitem[1984]{sellwood_carlberg_1984}
Sellwood J.A., Carlberg R.G., 1984, {\em Spiral instabilities provoked by accretion and star formation}, ApJ 282, 61--74

\bibitem[2002]{sepkoski_2002}
Sepkoski J., 2002, {\em A compendium of fossil marine animal genera}, (eds Jablonski D.\ \& Foote M.), Bull.\ Am.\ Paleontology, no.\ 363 (Paleontological Research Institution, Ithaca)

\bibitem[1996]{sivia_1996}
Sivia D.S., 1996, {\em Data analysis: A Bayesian tutorial}, Oxford University Press

\bibitem[2007]{smith_2007}
Smith A.B., 2007, {\em Marine diversity through the Phanerozoic: problems and prospects}, Journal of the Geological Society 164, 731--745

\bibitem[2005]{smith_mcgowan_2005}
Smith A.B., McGowan A.B., 2005, {\em Cyclicity in the fossil record mirrors rock outcrop area}, Biology Letters 1, 443--445

\bibitem[2008]{sober_2008}
Sober E., 2008, {\em Evidence and evolution}, Cambridge University Press

\bibitem[2003]{shaviv_2003}
Shaviv N.J., 2003, {\em The spiral structure of the Milky Way, cosmic rays, and ice age epochs on Earth}, New Astronomy 8, 39--77

\bibitem[2005]{shaviv_2005}
Shaviv N.J., 2005, {\em On climate response to changes in the cosmic ray flux and radiative budget}, J.\ Geophysical Research 110, A08105, 1--15

\bibitem[2003]{shaviv_veizer_2003}
Shaviv N.J., Veizer J., 2003, {\em Celestial driver of Phanerozoic climate?}, July 2003, 4--10

\bibitem[1983]{shoemaker_1983}
Shoemaker E.M., 1983, {\em Asteroid and comet bombardment of the Earth}, Ann.\ Rev.\ Earth Planet.\ Science 11, 461--494

\bibitem[1986]{shuter_klatt_1986} 
Shuter W.L.H., Klatt C., 1986, {\em Phase modulation of the Sun's z oscillations}, ApJ 301, 471--477

\bibitem[2008]{sloan_wolfendale_2008}
Sloan T., Wolfendale A.W., 2008, {\em Testing the proposed causal link between cosmic rays and cloud cover}, Environ.\ Res.\ Letters 3, 1--6

\bibitem[1990]{stanley_1990}
Stanley S.M., 1990, {\em Delayed recovery and the spacing of major extinctions}, Paleobiology 16, 401--414

\bibitem[1978]{stellingwerf_1978}
Stellingwerf  R.F., 1978, {\em Period determination using phase dispersion minimization}, ApJ 224, 953--960

\bibitem[1985]{stigler_1985}
Stigler S.M., 1985, {\em Terrestrial mass extinctions and galactic plane crossings}, Nature 313, 159

\bibitem[1987]{stigler_wagner_1987}
Stigler S.M., Wagner M.J., 1987, {\em A substantial bias in nonparametric tests for periodicity in geophysical data}, Science 238, 940--945

\bibitem[1988]{stigler_wagner_1988}
Stigler S.M., Wagner M.J., 1988, Reply to Raup \& Sepkoski~\citep{raup_sepkoski_1988}, Science 241, 96--99

\bibitem[1979]{stothers_1979}
Stothers R., 1979, {\em Solar activity during classical antiquity}, A\&A 77, 121--127

\bibitem[1986]{stothers_1986}
Stothers R., 1986, {\em Periodicity of the Earth's magnetic reversals}, Nature 322, 444--446

\bibitem[2008]{sturrock_2008}
Sturrock P.A., 2008, {\em A Bayesian approach to power spectrum significance estimation, with application to solar neutrino data}, arXiv:0809.0276v1

\bibitem[2006]{svensmark_2006}
Svensmark H., 2006, {\em Imprint of Galactic dynamics on Earth's climate}, Astron.\ Nachr.\ 9, 866--870

\bibitem[1997]{svensmark_friis_1997}
Svensmark H.,Friis-Christensen E., 1997, {\em Variation of cosmic ray flux and global cloud coverage -- a missing link in solar--climate relationships}, Journal of Atmospheric and Solar-Terrestrial Physics 59, 1225--1232

\bibitem[2006]{tanaka_2006}
Tanaka H.K.M., 2006, {\em Possible terrestrial effects of a nearby supernova explosion -- Atmosphere's response}, J.\ Atmospheric and Solar--Terrestrial Physics 68, 1396--1400

\bibitem[2004]{teterev_etal_2004}
Teterev A.V., Nemtchinov I.V., Rudak L.V., 2004, {\em Impacts of large planetesimals on the early Earth}, Solar System Research 38, 43--52

\bibitem[2005]{thomas_etal_2005}
Thomas B.C., Melott A.L., Jackman, C.H., Laird C.M., Medvedev M.V., Stolarski R.S., Gehrels N., Cannizzo J.K., Hogan D.P., Ejzak L.M., 2005, {\em Gamma-ray bursts and the Earth: Exploration of atmospheric, biological, climatic, and biogeochemical effects}, ApJ 634, 509--533

\bibitem[1976]{thomson_1976} 
Thomson K.S., 1976, {\em Explanation of large sale extinctions of lower vertebrates}, Nature 261, 578--580

\bibitem[1997]{toon_etal_1997}
Toon O.B., Turco R.P., Covey C., 1997, {\em Environmental perturbations caused by the impacts of asteroids and comets}, Rev.\ Geophysics 35, 41--78

\bibitem[1989]{torbett_1989}
Torbett M.V., 1989, {\em Solar system and Galactic influences on the stability of the Earth}, Palaeogeography, Palaeoclimatology, Palaeoecology 75, 3--33

\bibitem[1984]{torbett_smo_1984}
Torbett M.V., Smoluchowski R., 1984, {\em Orbtial stability of the unseen solar companion linked to periodic extinction events}, Nature 311, 641--642

\bibitem[2005]{turon_etal_2005}
Turon C., O'Flaherty K.S. Perryman, M.A.C.\ (eds.), 2005, {\em The Three-Dimensional Universe with Gaia}, ESA, SP-576

\bibitem[2008]{vallee_2008}
Vall\'ee J.P., 2008, {\em New velocimetry and revised cartography of the spiral arms in the Milky Way -- A consistent symbiosis}, AJ 135, 1301--1310

\bibitem[1999]{veizer_etal_1999} 
Veizer J., Davin A., Azmy K., Bruckschen P., Buhl D., Bruhn F., Carden G.A.F., Diener A., Ebneth S., Godderis Y., Jasper T., Kortea C., Podlaha O.G., Strauss H., 1999, {\em $^{87}$Sr/\,$^{86}$Sr, $\delta^{13}$C and $\delta^{18}$O evolution of Phanerozoic seawater}, Chemical Geology 161, 58--99

\bibitem[2008]{wick_napier_2008} 
Wickramasinghe J.T., Napier W.M., 2008, {\em Impact cratering and the Oort cloud}, MNRAS 387, 153--157

\bibitem[1977]{wielen_1977}
Wielen R., 1977, {\em The diffusion of stellar orbits derived from the observed age-dependence of the velocity dispersion}, A\&A 60, 263--275

\bibitem[2005]{white_saunders_2005}
White R.V., Saunder A.D., 2005, {\em Volcanism, impact and mass extinctions: incredible or credible coincidences}, Lithos 79, 299--316

\bibitem[1984]{whitmire_jackson_1984}
Whitmire D.P. Jackson A.A., 1984, {\em Are periodic mass extinctions driven by a distant solar companion?}, Nature 308, 713--715

\bibitem[2001]{wignall_2001}
Wignall P.B., 2001, {\em Large igneous provinces and mass extinctions}, Earth Science Reviews 53, 1--33

%\end{verse}
%\end{flushleft}

\end{thebibliography}
\end{document}